\documentclass[pra,aps,reprint,amssymb,amsmath,nofootinbib,superscriptaddress]{revtex4-2}
\usepackage{graphicx}
\usepackage{physics}
\usepackage{suffix}
\usepackage{color}
\usepackage{microtype}
\usepackage[colorlinks=true,allcolors=blue]{hyperref}

\makeatletter
\DeclareFontFamily{OMX}{MnSymbolE}{}
\DeclareSymbolFont{MnLargeSymbols}{OMX}{MnSymbolE}{m}{n}
\SetSymbolFont{MnLargeSymbols}{bold}{OMX}{MnSymbolE}{b}{n}
\DeclareFontShape{OMX}{MnSymbolE}{m}{n}{
    <-6>  MnSymbolE5
   <6-7>  MnSymbolE6
   <7-8>  MnSymbolE7
   <8-9>  MnSymbolE8
   <9-10> MnSymbolE9
  <10-12> MnSymbolE10
  <12->   MnSymbolE12
}{}
\DeclareFontShape{OMX}{MnSymbolE}{b}{n}{
    <-6>  MnSymbolE-Bold5
   <6-7>  MnSymbolE-Bold6
   <7-8>  MnSymbolE-Bold7
   <8-9>  MnSymbolE-Bold8
   <9-10> MnSymbolE-Bold9
  <10-12> MnSymbolE-Bold10
  <12->   MnSymbolE-Bold12
}{}

\let\llangle\@undefined
\let\rrangle\@undefined
\DeclareMathDelimiter{\llangle}{\mathopen}{MnLargeSymbols}{'164}{MnLargeSymbols}{'164}
\DeclareMathDelimiter{\rrangle}{\mathclose}{MnLargeSymbols}{'171}{MnLargeSymbols}{'171}
\makeatother

\newcommand*{\kket}[1]{\left|#1\right\rrangle}
\WithSuffix{\newcommand*}\kket*[1]{|#1\rrangle}
\newcommand*{\bbra}[1]{\left\llangle #1\right|}
\WithSuffix{\newcommand*}\bbra*[1]{\llangle #1|}
\newcommand*{\bbraket}[2]{\left\llangle #1\middle|#2\right\rrangle}
\WithSuffix{\newcommand*}\bbraket*[2]{\llangle #1|#2\rrangle}

\def\equationautorefname#1#2\null{Eq.#1(#2\null)}

\newcommand{\sigm}{\hat{\sigma}}
\newcommand{\sigp}{\sigm^{\dagger}}
\newcommand{\Liouv}{\hat{\mathcal{L}}}

\begin{document}

\title[Long Title]{ Coherent multidimensional spectroscopy in polariton systems}

\author{Daniela Gallego-Valencia}
\email[]{daniela.gallegov@udea.edu.co}
\affiliation{Grupo de F\'{\i}sica At\'omica y Molecular, Instituto de F\'{\i}sica, Universidad de Antioquia, Medell\'{\i}n, Colombia}
\affiliation{Departamento de F\'{\i}sica Te\'orica de la Materia Condensada and Condensed Matter Physics Center (IFIMAC), Universidad
Aut\'onoma de Madrid, 28049 Madrid, Spain}

\author{Lars Mewes}
\email[]{lars.mewes@tum.de}
\affiliation{Technische Universit\"at M\"unchen, TUM School of Natural Sciences, Fakult\"at f\"ur Chemie, Professur f\"ur Dynamische Spektroskopien, Lichtenbergstra\ss e 4, 85748 Garching, Germany}

\author{Johannes Feist}
\email[]{johannes.feist@udea.edu.co}
\affiliation{Departamento de F\'{\i}sica Te\'orica de la Materia Condensada and Condensed Matter Physics Center (IFIMAC), Universidad
Aut\'onoma de Madrid, 28049 Madrid, Spain}

\author{Jos\'e Luis Sanz-Vicario}
\email[]{jose.sanz@udea.edu.co}
\affiliation{Grupo de F\'{\i}sica At\'omica y Molecular, Instituto de F\'{\i}sica, Universidad de Antioquia, Medell\'{\i}n, Colombia}

\date{\today}

\begin{abstract}
The fast dynamics of molecular polaritonics is scrutinized theoretically through the implementation of two-dimensional spectroscopy protocols.
We derive conceptually simple and computationally efficient formulas to calculate two-dimensional spectra for molecules,
each of them modeled as a system of two electronic states including vibrational relaxation, immersed in an optical cavity, thus coupled to quantized radiation. Cavity photon losses and molecular relaxation are incorporated into the Hamiltonian dynamics to form an open quantum system that is solved through a master equation.
In the collective case, the relaxation dynamics into dark states reveals to be the crucial factor to explain the asymmetries in both the diagonal and cross peaks of two-dimensional spectra
for long waiting times between excitation and detection, a feature shown by recent experiments. Our theoretical method provides a deeper insight in those processes that yield relevant signals in multidimensional molecular spectroscopy.
\end{abstract}

\maketitle

\section{Introduction}

Aggregates of organic molecules subject to confined electromagnetic fields in
extended cavities provide a testbed to understand the interplay between
excitons, vibrations and phonons, and photons in physics. Excitations in these
collective systems produce entangled quasiparticles (polaritons) that inherit
properties from both matter and light~\cite{herrera2016, ribeiro2018, feist2018,
fregoni2022}. The fast inner workings of these complex systems in the short time
domain are still far from being fully understood. To unveil the polariton
dynamics of these systems, ultrafast laser techniques based on pump-probe
principles can be applied. Coherent multidimensional spectroscopy (CMDS)
provides insight into the vibronic structure and ultrafast dynamics of molecular
systems upon optical excitation~\cite{mukamel2000, jonas2003, abramavicius2009,
fuller2015, saurabh2016, dorfman2018, tiwari2021}. CMDS yields correlated
signals between the frequencies of absorption and those of detection, which
allows for a deeper understanding of the fast excitation, emission and
relaxation mechanisms of the system. CMDS has already provided remarkable new
insights into photophysics and photochemistry~\cite{oliver2018, gelzinis2019two}
since it allows a direct spectroscopic observation of couplings, system-bath
interactions and energy relaxation within microscopic systems. Two-dimensional
infrared (2D-IR) spectroscopy and two-dimensional electronic spectroscopy (2DES)
are of particular relevance and revealed insightful details about polaritonic
systems under vibrational~\cite{ribeiro2018theory, xiang2018two, simpkins2023}
and electronic~\cite{Li20172DCoherent} strong coupling, respectively.

Model Hamiltonians in quantum optics have been quite successful to describe the
fundamental physics behind photon-matter interactions, from quantum Rabi to
Jaynes-Cummings (JC) models in the case of a single emitter~\cite{Jaynes1963,
Shore1993}, to Dicke and Tavis-Cummings (TC) models for a collection of
emitters~\cite{Dicke1954, Tavis1968, garraway2011}.

In this article, we theoretically study and analyze CMDS of organic molecule
polaritons under electronic strong coupling. 
For the sake of completeness and
clarity, we first study the prototypical JC case of a single emitter and then
extend the description to the TC case with multiple molecules.
Inspired by the asymmetric signals observed in recent CMDS experiments involving
molecular J-aggregates immersed in optical cavities~\cite{mewes2020}, we study
the 2DES signals derived from an open quantum system consisting of a TC
Hamiltonian with $N$ identical emitters, each described by two electronic states
subject to dissipative and relaxation processes induced by vibrational modes.
Our model is expected to apply to an ensemble of dye molecules with a weak exciton-phonon coupling characterized by small Huang-Rhys factors and Stokes shifts, so that the coupling to molecular vibrations can be treated perturbatively. In particular,
molecular J-aggregates form delocalized collective electronic states, whose coupling to vibrational modes is weak~\cite{Kasha1963, Hestand2018}. The polariton photodynamics of a TC model has also been investigated recently in connection with pump-probe spectroscopies~\cite{fassioli2021}. Other recent works on 2DES using quantum stochastic Liouville equation~\cite{Zhang2019,Mondal2023} or the Heisenberg-Langevin model~\cite{Zhang2023} are of interest.


The paper is organized as follows. The theoretical framework is explained in
\autoref{sec:Theory} where we review some aspects of the polariton structure of
JC and TC models relevant for our study. Since we deal with an open quantum
system, the master equation is solved using the Liouvillian superoperator. We
describe the representation of the Liouvillian matrix and the distribution of
its complex eigenvalues and give a comprehensive account of the routes to
calculate linear (absorption and emission) and especially the non-linear
multidimensional spectroscopic signals, arriving at a remarkably simple formula
expressing these signals using the eigenstate of the Liouvillian.

Our results are presented in \autoref{sec:Results}; first the linear spectra,
which already show asymmetry in the absorption and emission of the lower and
upper electronic polaritons. We give a detailed survey on the successive steps
that build up the asymmetries in 2DS, the separate role of populations and
coherences in the non-linear spectra, the partial components contributing to the
2DS according to the Feynman paths and the changes in the spectra as the number
of molecular emitters increases. We finally use our method to reproduce the main
features of 2DS obtained in recent experiments with molecular J-aggregates
within optical microcavities. In \autoref{sec:Conclusions} we present our
conclusions and some future perspectives.

\section{Theoretical Framework}
\label{sec:Theory}

\subsection{Hamiltonian and Master Equations}
The Tavis-Cummings model for $N$ identical two-level emitters interacting with a
single cavity mode is described by the following Hamiltonian (with the case
$N=1$ corresponding to the Jaynes-Cummings model)
\begin{equation}
\hat{H} = \hbar \omega_c \hat{a}^\dag \hat{a} + \hbar \omega_0 \sum_{i=1}^N \sigp_i \sigm_i +
\hbar g \sum_{i=1}^N (\hat{a}^\dag \sigm_i + \sigp_i \hat{a}),
\end{equation}
in terms of the photon creation (annihilation) operator $\hat{a}^\dag$
($\hat{a}$) and the excitation (deexcitation) operator $\sigp_i$ ($\sigm_i$) of
molecule $i$. The emitters with natural frequency $\omega_0$ interact with a
single mode cavity radiation with frequency $\omega_c$ (with a detuning
$\Delta=\omega_c-\omega_0$) that leads to an energy splitting $\Omega_\text{R} =
\sqrt{4N g^2 + \Delta^2}$ (called the Rabi splitting in the case $\Delta=0$).
When comparing results with different numbers of emitters $N$ below, we scale
$g$ so as to maintain $N g^2$ and thus $\Omega_\text{R}$ constant.


\begin{figure}[t]
\includegraphics[width=\linewidth]{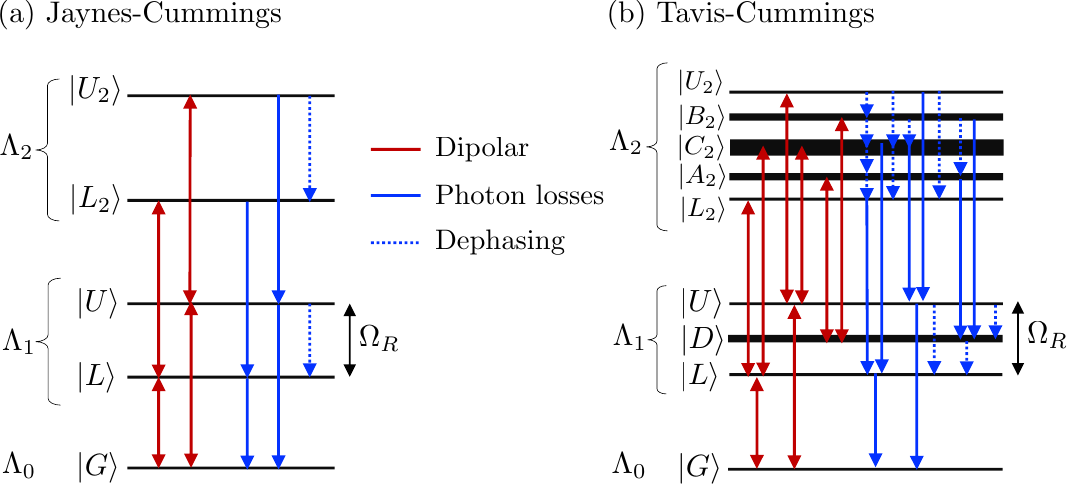}
\caption{ \label{fig:Figure1}
Scheme of energy levels and dominant transitions for (a) the Jaynes-Cummings
Hamiltonian with one emitter and (b) the Tavis-Cummings Hamiltonian with $N>2$
emitters.}
\end{figure}

Two-dimensional spectroscopy within a perturbative regime involves states in the
lowest excitation manifolds $\Lambda_0$, $\Lambda_1$ and $\Lambda_2$, where the
subscript denotes the number of excitations in the system. A general diagram of
the polariton energy levels of the JC Hamiltonian for one emitter and TC
Hamiltonian for $N>2$ emitters in resonance, $\omega_c=\omega_0$, is shown in
\autoref{fig:Figure1}. In both cases, the zero-excitation manifold contains a
single state, the ground state $\ket*{G}$. For the JC Hamiltonian,
in \autoref{fig:Figure1}(a), all other excitation manifolds $\Lambda_n$ contain two states, a
lower $\ket*{L_n}$ and an upper $\ket*{U_n}$ polariton with Rabi splitting
$\omega_{U_nL_n} \equiv \omega_{U_n}-\omega_{L_n} = \sqrt{n}\Omega_R$ (for simplicity, we drop the $n$ subindices for $n=1$) that increases with excitation
number $n$ (commonly called the vacuum Rabi splitting for $n=1$). The dominant
radiative transitions for external driving of the system are indicated with red
arrows in the figure. Blue arrows indicate the transitions induced by
dissipative processes, with cavity photon losses indicated by solid lines (these
are the same transitions that are accessible by external driving, i.e., the
solid blue and red arrows are the same) and vibrational relaxation between polaritons mediated by molecular dephasing-type interactions
indicated by dashed lines (which only occur between states within a given
excitation manifold). In \autoref{fig:Figure1}(b), we show a scheme of energy levels for a TC
Hamiltonian with $N$ emitters. The first excitation manifold $\Lambda_1$
consists of lower $\ket*{L}$ and upper $\ket*{U}$ states, separated by the Rabi
frequency $\Omega_R$ and $N-1$ degenerate dark states $\ket*{D}$. The second
excitation manifold $\Lambda_2$ has four energy levels: lower $\ket*{L_2}$ and
upper $\ket*{U_2}$ polaritons plus $N-1$ degenerate states $\ket*{A_2}$ and
$N-1$ degenerate states $\ket*{B_2}$, along with $N(N-3)/2 +1$ degenerate states
$\ket*{C}$. For the case
$N=2$, the states $A_2$ and $B_2$ are degenerate and the states $C_2$ disappear. The definitions for the vertical arrows are identical to panel (a). In the
limit $N\to\infty$, the TC model becomes linear, i.e., the collection of
identical emitters behaves like a harmonic oscillator~\cite{garraway2011}, and
the system is then described by normal modes (independent harmonic
oscillators)~\cite{Sanchez-Barquilla2022Perspective}. The $n$-excitation energy
levels can then be understood as corresponding to the excitation of $n$
independent quasiparticles (e.g., for $n=2$: two lower polaritons, or two upper
polaritons, or one lower polariton and one dark exciton, etc). For $N\gg n$,
this is still a useful picture even when $N$ is finite, with small energy shifts
that can be interpreted as interactions between the quasiparticles. 

We assume that the polariton system is embedded within an environment that
produces incoherent dynamics, in particular photon losses from the cavity. We
also treat the internal vibrational modes of the molecules (and their coupling
to vibrations and phonons in the host material) as an effective bath with a
dephasing-type interaction. The open quantum system is then treated under the
Markovian approximation~\cite{breuer2002, jeske2013, manzano2020}. Thus the
dynamics of the system density matrix is governed by a Liouville master equation
$\dot{\hat{\rho}}(t) = \Liouv [ \hat {\rho} (t) ] $, where the
Liouvillian superoperator is given by
\begin{equation}
\label{eq:LiouvilleEq}
\Liouv [ \hat {\rho} (t) ] 
= -\frac{i}{\hbar} [\hat{H}, \hat{\rho}(t) ] + \kappa \hat{L}_{\hat{a}} [ \hat{\rho}(y) ] + \sum_{i} \hat{\Gamma}_{\sigp_i \sigm_i} [ \hat{\rho}(t)],
\end{equation}
with a standard Lindblad term $\hat{L}_{\hat{a}}$ for the cavity losses with
decay rate $\kappa$ (lifetime $1/\kappa$)
\begin{equation}
\label{eq:Lindblad}
\hat{L}_{\hat{a}} [ \hat{\rho} (t)] = \hat{a} \hat{\rho}(t) \hat{a}^\dag - \frac{1}{2} \left\{ \hat{a}^\dag \hat{a}, \hat{\rho}(t) \right\}.
\end{equation}
The vibrational bath is described through a Markovian Bloch-Redfield-Wangsness
(BRW) superoperator $\hat{\Gamma}_{\hat{O}}$ for a single emitter in the form 
\begin{align}
\label{eq:BRW}
 & \hat{\Gamma}_{\hat{O} } [\hat{\rho } (t) ]  = \nonumber \\
 & -\frac{1}{\hbar^2} \sum_{m,n} \left\{ \hat{O}_{mn} \left[\hat{O}, |m \rangle \langle n | \hat{\rho}(t) \right] S_B (\omega_{mn})  + h.c.  \right\}
\end{align}
expressed in the Hamiltonian eigenbasis $\{ | n\rangle \}$ ~\cite{breuer2002,Jeske2015Bloch}. It is worth noting that the usual secular approximation is not implemented in this work while (Lamb) energy shifts are neglected. The BRW formalism describes the system-bath interaction induced by the molecular dephasing-type operators $\sigp_i \sigm_i$, which mediate vibrational relaxation between the polaritons~\cite{delPino2015}. The bath (assumed
independent for each molecule) is characterized by the noise power spectrum
\begin{equation}
  S_B(\omega) = \begin{cases} (1+n(\omega)) J(\omega) & \omega \geq 0 \\ n(-\omega) J(-\omega) & \omega < 0 \end{cases},
\end{equation} 
where $n(\omega) = 1/(e^{\hbar \omega/k_B T} - 1)$ is the Bose-Einstein thermal
population and $J(\omega)$ the spectral density of the bath. In the following,
we treat two cases: (i) a simplified model at zero temperature where
$n(\omega)=0$ and the spectral density $J(\omega) = \gamma$ is constant, and
(ii) a more realistic model at finite temperature with a spectral density of
Debye form, $J(\omega) = 2 \gamma \delta \omega/ (\omega^2 + \delta^2)$, with a molecular relaxation rate $\gamma$ and a cutoff parameter $\delta$.

\subsection{Liouvillian eigenvalues and eigenstates}

\label{sec:Liouvillecomplex}

\begin{figure}[t]
\includegraphics[width=0.75\linewidth]{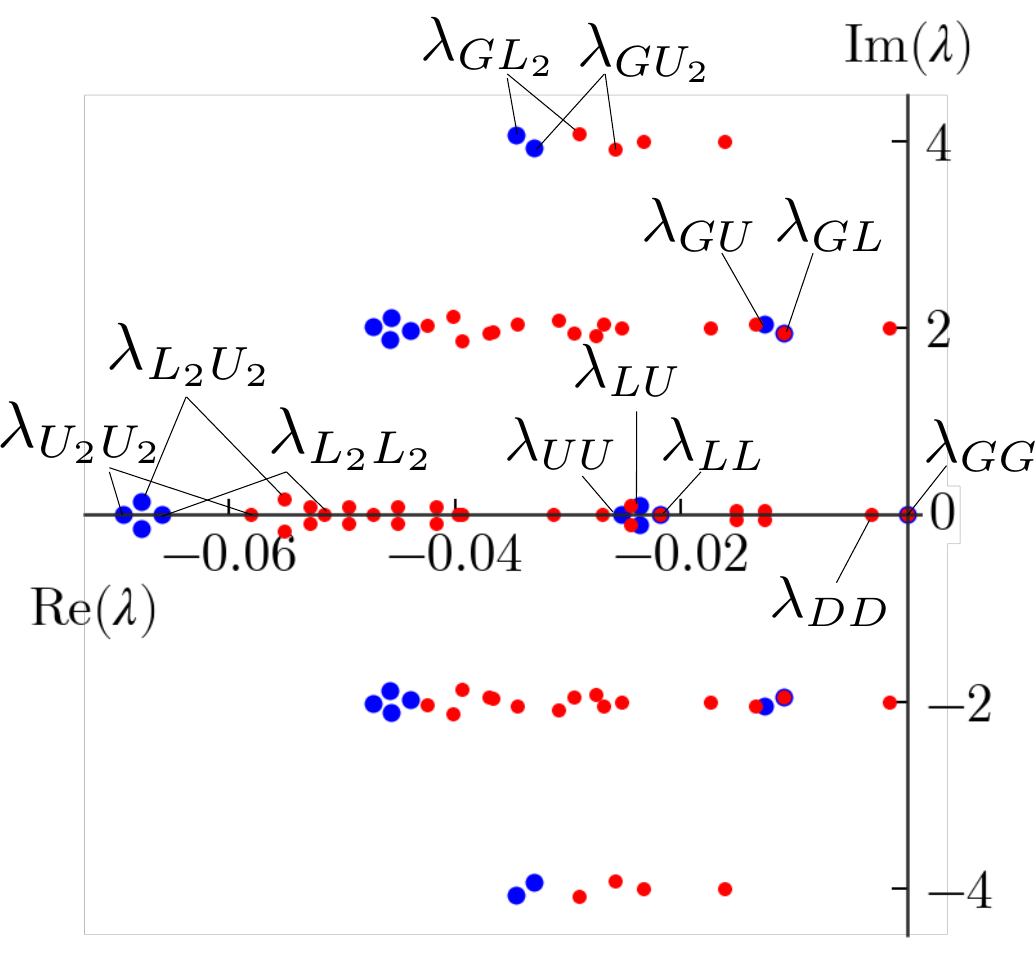}
\caption{ \label{fig:Figure2}
Location of the Liouvillian eigenvalues $\lambda_{\alpha \beta}= -\Gamma_{\alpha \beta} - i \omega_{\alpha \beta}$ in the complex plane for the open Jaynes-Cummings model (blue dots) and the open Tavis-Cummings model for $N=2$
emitters (red dots). In both cases, $\omega_c=\omega_e= 2$ eV and $\hbar \Omega_R$ = 0.1 eV. Cavity lifetime is 15 fs ($ \hbar \kappa$=44 meV) and molecular relaxation time is 50 fs ($ \hbar \gamma$=13 meV). The real part Re$[\lambda]$ is related to the total decay width (to all final channels) and the imaginary part Im$[\lambda]$ to energy differences between any two Hamiltonian eigenstates.
The most relevant eingenvalues corresponding to populations $\lambda_{\Lambda_i \Lambda_i}$ and coherences $\lambda_{\Lambda_i \Lambda_j}$
and the latter only when $\Lambda_1 \le \Lambda_2$ (for the opposite they are complex conjugated) are indicated in the plot.
}
\end{figure}

The master equation, \autoref{eq:LiouvilleEq}, describes the evolution of the
density matrix in the presence of dissipative terms. The Liouvillian
superoperator is not Hermitian and thus has complex eigenvalues. The density
operator can be represented in the Hilbert space basis, $\hat{\rho} =
\sum_{\alpha\beta} \rho_{\alpha\beta} \ketbra*{\alpha}{\beta}$, with a dimension
dim$_H^2$, where dim$_H$ is the dimension of the Hilbert space. The Liouvillian
superoperator is then an object with four indices, $\mathcal{L}_{\alpha \beta;
\alpha' \beta'}$, but can be interpreted as a matrix in the Liouville space
basis, whose elements $\kket*{\alpha\beta}$ map directly to the corresponding
Hilbert space operators $\ketbra*{\alpha}{\beta}$. The Liouvillian matrix
representation thus has dimension $\text{dim}_H^2 \times \text{dim}_H^2$. When it is diagonalizable, $\Liouv$ is
characterized by its (complex) eigenvalues $\lambda_i$ as well as its right and
left eigenvectors, $\kket*{v_i}$  and $\bbra*{v_i}$, with $\bbraket{v_i}{v_j} =
\delta_{ij}$. Often, the imaginary part of $\lambda_i$ is very close to an
energy difference between two eigenstates $\ket*{\alpha}$ and $\ket*{\beta}$ of
the Hamiltonian, and the corresponding eigenvalues of $\Liouv$ can be
approximately labelled as
\begin{equation}
\lambda_{\alpha \beta} = - \Gamma_{\alpha \beta} -i (\omega_{\alpha \beta} + \pi_{\alpha \beta}),
\end{equation}
where $\omega_{\alpha \beta} = (E_\alpha - E_\beta)/\hbar$ corresponds to the
energy difference between the two eigenstates and $\pi_{\alpha \beta}$ is a
small shift. For the case $\alpha=\beta$, both $\omega_{\alpha \beta}$ and
$\pi_{\alpha \beta}$ are zero and the Liouvillian eigenvalues are real. In
that case, the identification in terms of state labels can be performed by
inspection of the expansion coefficients of the Liouvillian eigenstates.

\begin{figure}[t]
\includegraphics[width=\linewidth]{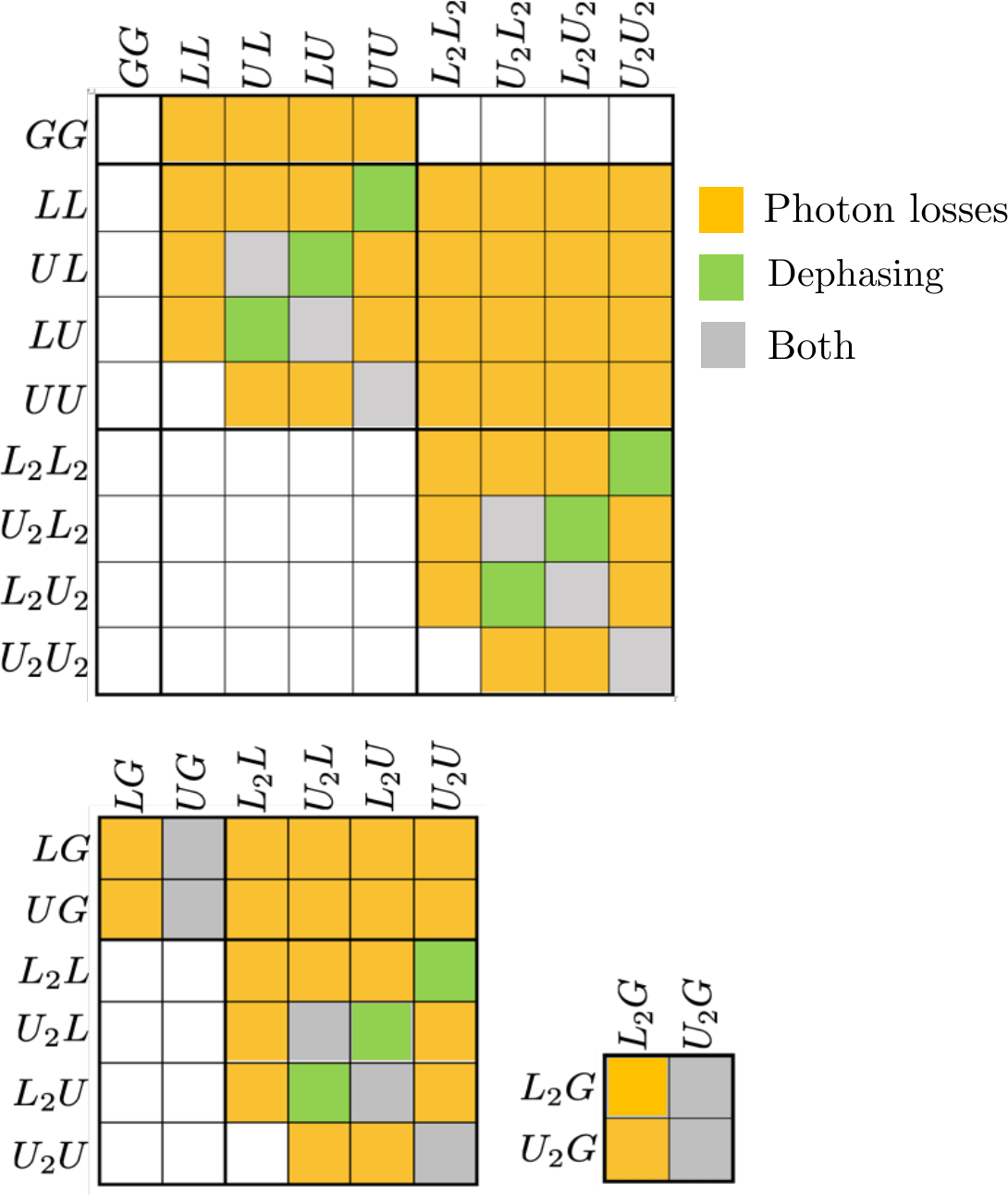}
\caption{ \label{fig:Figure3}
Scheme for the structure of the Liouvillian (25 x 25) matrix with elements
$\mathcal{L}_{\Lambda_i, \Lambda_j; \Lambda_k \Lambda_\ell}$ represented in
terms of the Hamiltonian basis set of the Jaynes-Cummings model up to the second
excitation manifold $\{G,L,U,L_2,U_2 \}$. The full matrix is organized as five
independent block matrices (the second 6 x 6 and third 2 x 2 block matrices in
the figure must be accompanied with their respective complex conjugates).
Colored grids correspond to non-zero matrix elements. Colors inside each grid
indicate the incoherent mechanism contributing to the element: photon loss
(orange), dephasing-type (green) or both (grey). }
\end{figure}

In \autoref{fig:Figure2}, we show the Liouvillian eigenvalues and corresponding
state labels (up to the second excitation manifold) of the JC model (5 states)
and the TC model for $N=2$ emitters (8 states), with 25 and 64 Liouvillian
eigenvalues, respectively. Eigenvalues corresponding to population dynamics,
$\lambda_{\alpha \alpha}$, lie along the real axis, while those corresponding to
coherences $\lambda_{\alpha \beta}$ have nonzero real parts. Notice that the
real parts of $\lambda$ for $N=1$ effectively act as a lower and upper bounds
for decay widths of polariton systems with $N > 1$ (except for the presence of
new dark states and their coherences), thus indicating in general a nonlinear
scaling of decays for higher excitation manifolds as $N$ increases. For the real
eigenvalues, one can observe a hierarchy in the total population decay rates,
$0=\Gamma_{GG} <$ $\Gamma_{LL}<$ $ \Gamma_{UU} <$ $\Gamma_{L_2 L_2} <$
$\Gamma_{U_2U_2}$. The fastest decay corresponds to the highest energy level
$\ket*{U_2}$ due to the large number of available decay channels (both photonic
and vibrational). It is worth noting that coherences also decay along with
populations.

The structure of the Liouvillian matrix (which is the generator of the
dissipative dynamics) also reflects the asymmetry in the different decay
processes. In \autoref{fig:Figure3}, we include a colored scheme of the
Liouvillian matrix for the JC model indicating those matrix elements responsible
for either photon losses or vibrational relaxation or both simultaneously. The matrix is
block diagonal, such that each block evolves in time independently. The first
matrix block contains elements with the form $(\Lambda_i \Lambda_i; \Lambda_j
\Lambda_j)$ for $i,j=0,1,2$ (note that labels $\Lambda_0 =\{G\}$, $\Lambda_1 =
\{L, U\}$, $\Lambda_2=\{L_2, U_2\}$ correspond to excitation manifolds). There
are transitions $\Lambda_i \Lambda_i \to \Lambda_{i-1} \Lambda_{i-1}$ due to
photon losses, while the opposite process (pumping) $\Lambda_i \Lambda_i \to
\Lambda_{i+1} \Lambda_{i+1}$ does not occur. Similarly, within the second block,
the photon losses show up in transitions between coherences $\Lambda_{2}
\Lambda_{1} \to \Lambda_{1} \Lambda_{0}$, but not in the opposite direction.
Molecular vibrational relaxation is present only within the blocks $(\Lambda_i \Lambda_i ;
\Lambda_i \Lambda_i)$ with $i=1,2$, $(\Lambda_{i+1} \Lambda_i ; \Lambda_{i+1}
\Lambda_i)$ with $i=0,1$ and $(\Lambda_{i+2} \Lambda_i; \Lambda_{i+2}
\Lambda_i)$ with $i=0$.

While the Liouvillian matrix elements may be calculated in closed form, the
dynamics $\hat{\rho}(t) = e^{\Liouv t} \hat{\rho}(0)$ must be solved
numerically. As an example, we discuss an initial state that is a superposition
of polaritons within the first excitation manifold, i.e., $\hat{\rho}(0) =
\ketbra*{\Psi}$ with $\ket*{\Psi} = C_L \ket*{L} + C_U \ket*{U}$. The evolution
is then fully determined by the zero- and one-excitation manifolds, and the
reduced Liouvillian master equation takes the form
\begin{widetext}
 \begin{equation}
  \dv{t}
 \begin{pmatrix}
  \rho_{GG} (t) \\
  \rho_{LL} (t) \\
  \rho_{UL} (t) \\
  \rho_{LU} (t) \\
  \rho_{UU} (t)
 \end{pmatrix}
=
 \begin{pmatrix}
 0 & \kappa/2 & \kappa/2 & \kappa/2 & \kappa/2 \\
 0 & -\kappa/2 & -\kappa/4 & -\kappa/4 & \gamma/4 \\
 0 & -\kappa/4 & -\kappa/2-\gamma/8-i\Omega_R & \gamma/8 & -\kappa/4 \\
 0 & -\kappa/4 & \gamma/8 & -\kappa/2 - \gamma/8 + i \Omega_R & -\kappa/4 \\
 0 & 0 & -\kappa/4 & -\kappa/4 & -\kappa/2-\gamma/4
\end{pmatrix}
 \begin{pmatrix}
 \rho_{GG} (t) \\
 \rho_{LL} (t) \\
 \rho_{UL} (t) \\
 \rho_{LU} (t) \\
 \rho_{UU} (t)
 \end{pmatrix}.
\end{equation}
\end{widetext}
Of the five eigenvalues of this Liouvillian matrix, two have simple analytic
forms, $\lambda_{GG} = 0$, $\lambda_{UU} = -\kappa/2 - \gamma/4$, while the
remaining three, $\lambda_{LL}$, $\lambda_{LU}$, $\lambda_{UL}$, are the
analytic solutions of the polynomial $256 \kappa \Omega_R^2 + (16 \gamma \kappa
+ 32 \kappa^2 + 64\omega_R^2)\lambda + (2\gamma + 12 \kappa) \lambda^2 +
\lambda^3 = 0$. These solutions can be approximated as $\lambda_{LL} \approx
-\kappa/2$, $\lambda_{LU} \approx -\kappa/2 - \gamma/8 + i \tilde{\Omega}_R$ and
$\lambda_{UL} \approx -\kappa/2 - \gamma/8 - i \tilde {\Omega}_R$, with a
shifted Rabi frequency $\tilde{\Omega}_R = \omega_{UL} + \pi_{UL}$ (see also
\autoref{fig:Figure2} to locate these relevant eigenvalues). Time propagation is
straightforward within the diagonalized (spectral) representation of the
Liouvillian, with populations given by $\rho_{\alpha\alpha}(t) = \sum_{\alpha'
\beta'} C^{\alpha \alpha}_{\alpha' \beta'} e^{\lambda_{\alpha' \beta'} t}$,
where $C^{\alpha \alpha}_{\alpha' \beta '} = \bbraket*{\alpha \alpha}{v_{\alpha'
\beta'}} \bbraket*{v_{\alpha' \beta'}}{\hat{\rho}(0) }$. In particular, the
populations for the above-mentioned initial condition can be written as
\begin{multline}
\label{eq:lambdapop}
\rho_{\alpha \alpha}(t) = C^{\alpha \alpha}_{LL} e^{-\Gamma_{LL} t} + C^{\alpha \alpha}_{UU} e^{-\Gamma_{UU} t} \\
+ 2 e^{-\Gamma_{UL} t}\Re\left[C^{\alpha\alpha}_{UL} e^{i \tilde{\Omega}_R t}\right],
\end{multline}
where we have used that $C^{\alpha \alpha}_{UL} = (C^{\alpha \alpha}_{LU})^*$.
\autoref{eq:lambdapop} involves two terms with exponential decays for
Liouvillian eigenstates associated to populations, for the $\kket*{v_{LL}}$ and
$\kket*{v_{UU}}$ components, and additional damped oscillatory terms due to the
Liouvillian coherences $\kket*{v_{LU}}$ and $\kket*{v_{UL}}$. As shown below, this result
qualitatively explains the behavior of the diagonal and cross peaks in 2D
spectra after excitation, with respect to the waiting time delay; namely,
exponential decay accompanied by Rabi oscillations.

The TC model with $N=2$ emitters adds a new dark state $D$ in the dynamics and
the master equation now involves a 6 x 6 reduced Liouvillian matrix as shown in
\autoref{eq:TCLiouville}. One realizes that while photon loss rates enter in TC
in analogy with the JC model, the vibrational relaxation rates enter differently and not only
in the new column and row involving the dark state. Thus, the presence of dark
states modifies the global dynamics of populations and coherences within the
whole system.
\begin{widetext}
 \begin{eqnarray}
 \label{eq:TCLiouville}
 \dv{t}
 \begin{pmatrix}
 \rho_{GG} (t) \\
 \rho_{LL} (t) \\
 \rho_{UL} (t) \\
 \rho_{DD} (t) \\
 \rho_{LU} (t) \\
 \rho_{UU} (t)
 \end{pmatrix}
 =
 \begin{pmatrix}
 0 & \kappa/2 & \kappa/2 & 0 & \kappa/2 & \kappa/2 \\
 0 & -\kappa/2 & -\kappa/4 + \gamma/8 & \gamma/4 & - \kappa/4 + \gamma/8 & \gamma/8 \\
 0 & -\kappa/4 & -\kappa/2-3\gamma/16-i\Omega_R & -\gamma/8 & \gamma/16 & -\kappa/4 + \gamma/8 \\
 0 & 0 & -\gamma/8 & -\gamma/4 & -\gamma/8 & \gamma/4 \\
 0 & -\kappa/4 & \gamma/16 & -\gamma/8 & -\kappa/2 - \gamma/8 + i \Omega_R & -\kappa/4 + \gamma/8 \\
 0 & 0 & -\kappa/4 & 0 & -\kappa/4 & -\kappa/2-3\gamma/8
\end{pmatrix}
\begin{pmatrix}
 \rho_{GG} (t) \\
 \rho_{LL} (t) \\
 \rho_{UL} (t) \\
 \rho_{DD} (t) \\
 \rho_{LU} (t) \\
 \rho_{UU} (t)
 \end{pmatrix}
\end{eqnarray}
\end{widetext}
The Liouvillian real eigenvalues of the JC model corresponding to the decay widths $\Gamma$ have a dependence upon the choice of rate parameters $\kappa$ (photon loss) and $\gamma$
(vibrational relaxation). The decay widths show a linear scaling with $\kappa$, with a slope $\frac{2 n -1}{2}$ for each excitation manifold $\Lambda_n$ and remains with the same trend for any number
of molecular emitters $N$. Widths also display a linear scaling against the relaxation rate $\gamma$ with slopes that depend on the nature of the state more than on the excitation manifold.
For instance, in the JC model, $\lambda_{LL}$ and $\lambda_{L_2 L_2}$ share zero, $\lambda_{UU}$ and $\lambda_{U_2 U_2}$ share the same $\gamma/4$ and coherences $\lambda_{LU}$
and $\lambda_{UL}$ have $\gamma/8$.
However, the molecular relaxation has a more complex scaling rule when $N$ increases.

\subsection{Linear Absorption and Emission Spectra}

Absorption and linear emission spectroscopy characterize the linear response of
the system upon excitation. For absorption, the spectra is obtained as the
Fourier transform of the two-time correlation function
\begin{equation}
S_{\text{Abs}} (\omega_L) = \Re\left[ \int_{0}^{\infty} \expval*{\hat{a} (t') \hat{a}^\dag (0)} e^{i \omega_L t'} \dd{t'} \right].
\end{equation}
Note that in Fabry-Perot-like cavities the laser pumping through the mirrors is treated following the input-output theory, where the external laser pulses only drive the cavity mode(s) directly, which in turn mediate the coupling to the molecules~\cite{Lentrodt2020}.
Emission is obtained from the response under excitation by a weak
continuous-wave laser driving field with frequency $\omega_L$, i.e.,
$\hat{H}_L(t) = E_0 ( \hat{a} e ^{i\omega_L t} + \hat{a}^\dag e^{-i \omega_L
t})$, described within the rotating-wave approximation. Transforming to a
rotating frame with the unitary operator $U(t)=e^{i \hat{a}^\dag \hat{a}
\omega_L t/\hbar}$ yields a time-independent Hamiltonian without affecting the
other terms in the Liouvillian superoperator. The emission spectra can then be
calculated from another two-time correlation function in the steady state $s$
(fulfilling $\dot{\rho}$ = 0), i.e.,
\begin{equation}
S_{\text{Em}} (\omega_L, \omega) = \Re \left[ \int_0^\infty \expval*{\hat{a}^\dag (t') \hat{a}(0)}_{s} e^{-i \omega t'} \dd{t'} \right].
\end{equation}
Note that the correlation does not disappear for $t\to\infty$, but reaches a
constant value, $\lim_{t\to\infty}\expval*{\hat{a}^\dag (t) \hat{a}(0)}_{s} =
|\!\expval*{\hat{a}}_{s}\!|^2$ corresponding to elastic scattering of the laser,
i.e., a delta peak at frequency $\omega_L$. This contribution is not plotted in
the emission spectra shown below.

\subsection{Non-linear Two-dimensional Spectroscopy}

\begin{figure*}[tbp]
\includegraphics[width=0.75\linewidth]{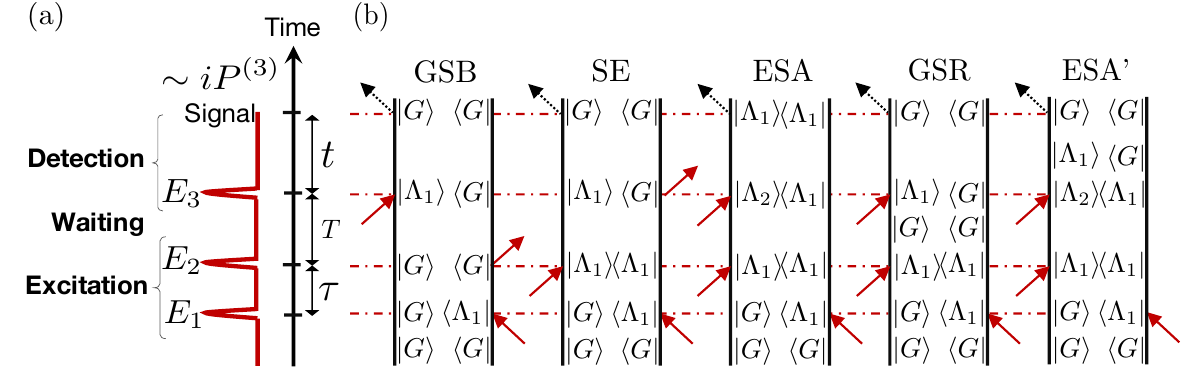}
\caption{ \label{fig:Figure4}
(a) 2DS scheme showing three delayed laser pulses separated by coherence time,
$\tau$, and waiting time $T$. The signal is emitted during the detection time
$t$. (b) Set of double sided Feynman diagrams for the rephasing phase-matching
condition ($-k_1+k_2+k_3$) considering the excitation manifolds $\{\Lambda_0,
\Lambda_1$, $\Lambda_2 \}$, that correspond to GSB, SE and ESA processes, plus
those GSR and ESA$^{\prime}$ processes derived when relaxation occurs after
excitation. Red arrows indicate dipole interactions with the laser pulses and
outgoing black dotted arrows represent the detected signal. Note that the notation $|\Lambda_1\rangle \langle \Lambda_1|$ indicates both populations and coherences, since $\Lambda_1 = {L,U}$}.
\end{figure*}

While linear spectroscopies only involve the two lowest excitation manifolds
$(\Lambda_0, \Lambda_1)$, 2D spectroscopy to lowest order involves the three
lowest excitation manifolds $(\Lambda_0, \Lambda_1, \Lambda_2)$, and can be
understood as a coherent excitation energy-resolved pump-probe experiment. As
depicted in \autoref{fig:Figure4}(a), the first interaction of the system with
the field $E_1(t)$ creates a coherence that evolves according to the Liouvillian
during the coherence time delay $\tau$. The interaction with $E_2(t)$ converts
the initial coherence into a population on the ground or excited-state. The
system evolves during the waiting time $T$ before a third field interaction with
$E_3(t)$ generates a coherence that eventually radiates a signal field during
the detection time period $t$, which is proportional to the third-order
polarization $P^{(3)}(\tau, T, t)$.


Two common routes to compute third-order spectra are to either directly use
third-order perturbation theory~\cite{mukamel1999, cho2009, hamm2011}, or to
extract the third-order component of the density matrix from the full
non-perturbative solution of the quantum dynamic equations~\cite{gelin2009,
deSio2019}. We follow the first route here, which provides conceptually simple
way to understand the underlying physics. The third-order polarizability is thus
given by $P^{(3)} (t) = \Tr[\hat{\mu} \; \rho^{(3)}]$, where $\rho^{(3)}$ corresponds to the
third-order perturbative component of the density matrix, and $\hat{\mu}$ is the
operator coupling the system to the incoming laser. 

$\rho^{(3)}$ can be expressed via time-ordered integrals, where the initial
density operator $\rho_0$ is subject to three laser interactions at
the times $t_1 < t_2 < t_3$:
\begin{widetext}
\begin{equation}
 \rho^{(3)}(t) = \int_{t_0}^t \dd{t_3} \int_{t_0}^{t_3} \dd{t_2} \int_{t_0}^{t_2} \dd{t_1}
\mathcal{G}(t-t_3) \breve{V}(t_3) \mathcal{G}(t_3 - t_2) \breve{V}(t_2) \mathcal{G}(t_2 - t_1) \breve{V}(t_1) \rho(t_0).
\end{equation}
\end{widetext}
Here, the notation $\breve{V}(t)$ indicates a superoperator that applies the interaction operator $\hat{V}(t) = -\hat{\mu} E(t)$ as a commutator,
$\breve{V}(t) \rho(t) = \frac{1}{i\hbar}[\hat{V}(t), \rho(t)]$, with $E(t)$ the driving laser
field amplitude, while $\mathcal{G}(\Delta t) = e^{\mathcal{L} \Delta t}$ indicates the field-free time
propagator (Green's function) over a time interval $\Delta t$. The third-order polarization can be reexpressed in terms of the third-order response function $S(t,T,\tau)$ (being $t, \ T,$ and $\tau$ general time intervals) as
\begin{align}
& P^{(3)}(t_f) = \int_{t_0}^{t_f} \dd{t_3} \int_{t_0}^{t_3} \dd{t_2} \int_{t_0}^{t_2} \dd{t_1} \nonumber\\
           &\qquad E(t_3) E(t_2) E(t_1) S(t_f-t_3, t_3-t_2, t_2-t_1),
\end{align}
where
\begin{align}
& S(t,T,\tau) = \Tr[ \hat{\mu} \mathcal{G}(t) \breve{\mu} \mathcal{G}(T) \breve{\mu} \mathcal{G}(\tau) \breve{\mu} \rho(t_0)].
\end{align}

We now assume the sudden impulsive limit for the laser fields involved, with a
Dirac delta as the envelope function for the three laser fields $E(t) = E^0
\delta(t - \tau_i) \exp[i (\pm k_i r \mp \omega_i t)]$, for $i=1,2,3$ (note that here $E^0$ has dimensions of electric field by time). This is a
reasonable approximation for sufficiently short non-overlapping laser pulses. It
implies a large spectral bandwidth and produces simplified expressions since it
gives $P^{(3)}(t_f) = S(t,T,\tau)$, where now $t= t_f-\tau_3$, $T = \tau_3-\tau_2$, and $\tau = \tau_2-\tau_1$. For target systems
that conserve momentum (as planar Fabry-Pérot cavities do in the in-plane
directions), phase-matching conditions imply that the signal can be split into
distinct components for which emission occurs into different directions that
can be distinguished experimentally~\cite{hamm2011}. The most typical choices
are the \emph{rephasing} (R) ($-k_1+ k_2 + k_3 $) and \emph{non-rephasing} (NR)
$(+k_1-k_2+k_3)$ components, since the sum of both contributions produces the
total absorptive 2DS\@. The chosen sign $+$ or $-$ for $k_i$ in the phase-matching
condition determines (within the rotating-wave approximation) whether the
excitation $\hat{\mu}^+$ or deexcitation $\hat{\mu}^-$ part of the coupling
operator acts for each interaction. For instance, the response function for the
rephasing matching condition reads
\begin{equation}
S_{R}(t,T,\tau) = \Tr\left[\hat{\mu}^- \mathcal{G}(t) \breve{\mu}^+ \mathcal{G}(T) \breve{\mu}^+ \mathcal{G}(\tau) \breve{\mu}^- \rho (t_0) \right].
\end{equation}
Here the last operator $\hat{\mu}^-$ corresponds to the signal emission. By expanding the nested commutators one arrives at eight terms, each of them
corresponding to a particular double-sided Feynman diagram of the rephasing
process. Under coherent evolution of the system and within the rotating-wave
approximation, only three out of eight diagrams contribute to the signal, namely
ground state bleaching (GSB), stimulated emission (SE) and excited state
absorption (ESA), as included in \autoref{fig:Figure4}(b). Taking into account
incoherent processes such as relaxation (which are often orders of magnitude
faster in polaritonic systems than in isolated molecules), the system does not
stay in the same eigenstate even during the field-free evolution and additional
Feynman diagrams become relevant. Note that the Feynman diagrams shown in
\autoref{fig:Figure4}(b) are based on labels corresponding to excitation
manifolds, not explicit states, such that relaxation processes that occur within
an excitation manifold are implicitly contained in each one. One additional
path, the ground state recovery (GSR), involves population relaxation from the
first excitation manifold, i.e., $\ket*{\Lambda_1} \bra*{\Lambda_1} \to \ket*{G}
\bra*{G}$, and ensures that the GSB signal decays when the molecular system
relaxes back into the ground state (see~\cite{valkunas2013}). Another path,
ESA$^\prime$, is equivalent to ESA but with an additional decay process
happening between excitation by the third pulse and photon emission $t$, i.e.,
$\ket*{\Lambda_2} \bra*{\Lambda_1} \to \ket*{\Lambda_1} \bra*{G}$. To our
knowledge, this pathway has not been discussed so far, but we find it relevant
in the build-up of the 2DS at short $T$. Note that both GSR and ESA$^\prime$ paths involve relaxation due to decay of cavity photons thus connecting different excitation manifolds. Vibrational relaxation only happens within the same excitation manifold. 

The rephasing 2D spectrum $S_{R} (\omega_t, T, \omega_{\tau})$ is obtained after
the 2D Fourier transform of $S_{R}(t,T,\tau)$, while the total absorptive 2D
spectrum is obtained by adding also the non-rephasing contribution,
$S_{\text{Abs}} (\omega_t, T, \omega_\tau)= \Re (S_R + S_{NR})$.

\subsection{Simple formula for 2DS}
\label{sec:Formula2DS}
Here we derive a simple formula for the computation of 2DS (in principle valid
for any open quantum system described by a diagonalizable Liouvillian). As
mentioned above, the density operator can be vectorized with a compact notation
based on a single combined index. With this vectorized form, the time-dependent
master equation reads $\dv{t} \kket*{\rho(t)} = \Liouv \kket*{\rho(t)}$. The
formal solution is simply $\kket*{\rho(t)} = e^{\Liouv t} \kket*{\rho(0)}$,
which can be straightforwardly represented within the basis of eigenstates
$\kket*{v_i}$ of the Liouvillian (with $\Liouv \kket*{v_i} = \lambda_i
\kket*{v_i}$) as $\kket*{\rho(t)} = \sum_i c^{(0)}_i e^{\lambda_i t}
\kket*{v_i}$, where $c_i^{(0)} = \bbraket*{v_i}{\rho(0)}$ are the coefficients
of the initial density matrix in the Liouvillian eigenbasis. The matrix
representation of the Liouvillian is neither Hermitian nor symmetric and the
left and right eigenvectors are thus distinct, but do fulfill
$\bbraket*{v_i}{v_j} = \delta_{ij}$.

Expressing the superoperator ($\breve{\mu} \hat{\rho} = \frac{1}{i\hbar} [\hat{\mu},
\hat{\rho}]$) in the eigenbasis of the field-free Liouvillian, $\breve{\mu}_{ij}
= \bbraket{v_i}{\breve{\mu} v_j}$, then leads to a compact expression for the
third-order density matrix. For instance, in the rephasing matching case, the
result is
\begin{equation}
\kket*{\rho{(t)}} = \sum_{ijkl} e^{\lambda_k t} e^{\lambda_j T} e^{\lambda_i \tau} \breve{\mu}^{+}_{kj} \breve{\mu}^{+}_{ji} \breve{\mu}^{-}_{il} c_l^{(0)} \kket*{v_k}.
\end{equation}

The field-free propagation during the times $(t,T,\tau)$ is reflected in the exponential factors.

The third order non-linear response function, $S = \Tr
[\hat{\mu}^{-} \rho(t)]$, is obtained in Liouville space simply as
$S =  \bbraket*{\hat{\mu}^{+}}{\rho(t)}$. Thus the expression for the rephasing response function becomes $S_R(\tau, T, t) = \sum_{ijkl}
e^{\lambda_k t} e^{\lambda_j T} e^{\lambda_i \tau} \hat{\mu}^{-}_{k}
\breve{\mu}^{+}_{kj} \breve{\mu}^{+}_{ji} \breve{\mu}^{-}_{il} c_l^{(0)}$, where $\hat{\mu}^{-}_{k} =
\bbraket*{\hat{\mu}^{+}}{v_k}$. Performing a 2D Fourier transform over $t$ and
$\tau$ and reorganizing the summations then leads to a remarkably simple
expression for the 2D spectrum:
\begin{equation}
\label{eq:spectra}
S(\omega_t, T, \omega_\tau) = \sum_j E_j(\omega_\tau) e^{\lambda_j T} D_j (\omega_t),
\end{equation}
with an excitation mask function $E_j = E_j^{NR} + E_j^R$, where
\begin{equation}
E_j^{NR/R}(\omega_{\tau}) = \sum_{il} \frac{1}{\pm i \omega_\tau + \lambda_i} \breve{\mu}^{\mp}_{ji} \breve{\mu}^{\pm}_{il} c_l^{(0)}
\end{equation}
(the rephasing signal is located in the $(\omega_t, -\omega_{\tau})$ quadrant~\cite{hamm2011}) and a detection mask function
\begin{equation}
D_j(\omega_t) = \sum_k \frac{1}{i \omega_t + \lambda_k} \hat{\mu}^{-}_{k} \breve{\mu}^{+}_{kj},
\end{equation}
that only depend on excitation $\omega_\tau$ and detection $\omega_t$
frequencies, respectively. Notice that the index $j$ in \autoref{eq:spectra}
runs over the whole set of eigenstates of the Liouvillian. However, in practice,
only a few of them contribute simultaneously to the absorption and emission
masks. We note that these mask functions are not directly observable quantities:
the complex excitation mask $E_j(\omega_\tau)$ does not in general coincide with
the linear absorption spectrum, and the complex detection mask $D_j(\omega_t)$
does not correspond to the emission spectrum.

\begin{figure}[t]
\includegraphics[width=\linewidth]{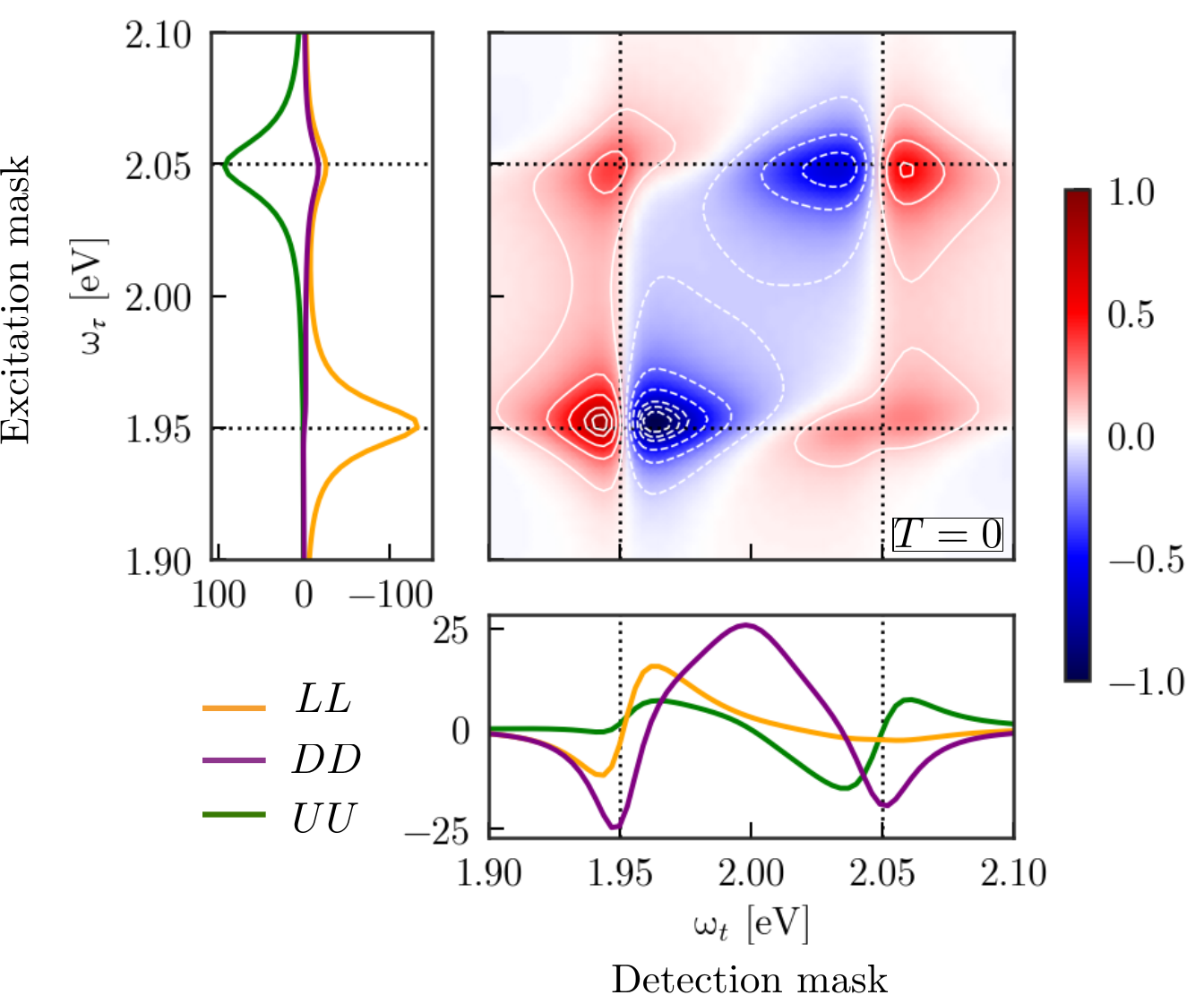}
\caption{ \label{fig:Figure5}
2D absorptive spectrum $S_{\text{Abs}}(\omega_t,T,\omega_\tau)$ for a TC model with $N=2$ molecules in resonance with the cavity $\omega_0=\omega_c=2$ eV, Rabi frequency $\Omega_R=0.1$ eV, cavity lifetime 15 fs and dephasing lifetime 50 fs, and BRW with a spectral function $J(\omega)=\gamma$ (at zero temperature)
 for waiting time $T=0$. The spectrum is built following \autoref{eq:spectra} with the index $j$ running only over the three leading Liouvillian eigenvalues, that we label here as $\{\lambda_{LL},
 \lambda_{UU}, \lambda_{DD}\}$
 (see \autoref{fig:Figure2}). The contributions to
 the excitation Re$[E_j(\omega_\tau) ]$ and detection Re$[D_j(\omega_t)]$ masks are also plotted along the corresponding axis. 2D spectrum is normalized to unity at its maximum value.
}
\end{figure}

An example of the construction of the 2DS through the excitation and detection
mask functions is included in \autoref{fig:Figure5} (here and in the following,
we set $\hbar=1$). This plot corresponds to $\Re[S(\omega_t, T, \omega_\tau)]$
for a TC model with $N=2$ molecular emitters with Rabi energy splitting
$\Omega_R=0.1$ eV subject to relaxation due to photon loss and dephasing, and
for a chosen waiting time $T=0$ so that the factor $e^{-\lambda_j T}$ is unity.
In this particular case we find that from the 64 complex eigenvalues to be
included in the sum in \autoref{eq:spectra} (see also
\autoref{sec:Liouvillecomplex}), in practice we only need three eigenvalues
(those corresponding to populations $\lambda_{LL}, \lambda_{UU}$ and
$\lambda_{DD}$). In this case, even though the excitation mask barely shows a
contribution from the dark state $\lambda_{DD}$, this becomes dominant in the
detection mask and is responsible for the asymmetric signals at the cross-peaks
$(\omega_\tau,\omega_t) = (1.95,2.05)$ eV and $(2.05,1.95)$ eV. In contrast,
this asymmetry between the two cross peaks is not present in the JC model
(without dark states) at $T=0$ (see \autoref{fig:Figure10} below).

\section{Results}
We have implemented the expressions above in a numerical code written in Python
and based on the QuTiP quantum optics toolbox~\cite{johansson2013}, used to
produce all results shown below. Due to the computational efficiency, all
results presented in this work can be calculated on a desktop computer.

\label{sec:Results}

\begin{figure}[tb]
\includegraphics[width=\linewidth]{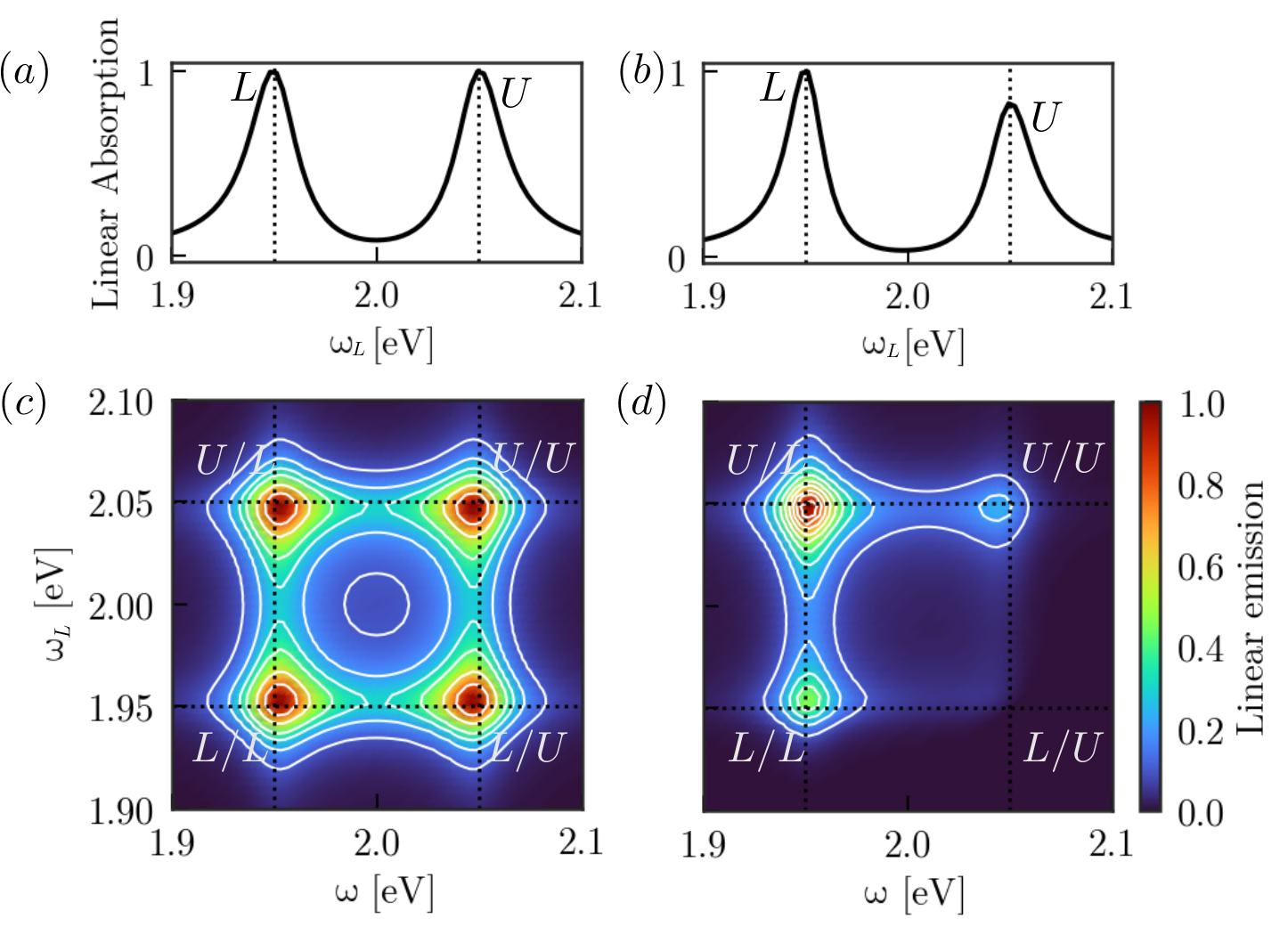}
\caption{\label{fig:Figure6}
(a) Linear absorption spectrum $S_{\text{Abs}} (\omega_L)$ for a polariton system with $N=2$ molecules within a resonant cavity $\omega_c=\omega_0=2$ eV with a Rabi splitting $\Omega_R=0.1$ eV and
cavity lifetime 15 fs and dephasing lifetime 50 fs. Absorption bands are located at $\omega_{LG}=1.95$ eV and $\omega_{UG}=2.05$ eV. Results obtained by using the Lindblad formalism for both cavity and molecular dephasing.
(b) Same as (a) but using Bloch-Redfield-Wangsness theory, only for the molecular dephasing.
(c) Excitation-emission spectrum $S_{\text{Em}} (\omega_L, \omega)$ for the same polariton system as a function of the laser excitation frequency $\omega_L$ and emission frequency $\omega$, using the Lindblad formalism.
(d) Same as (c) but BRW theory replaces Lindblad formalism for dephasing.
All spectra are normalized to unity at their respective maxima.
}
\end{figure}

\subsection{Linear Spectroscopy}

Upper (U) vs.\ lower (L) polariton asymmetries in the signal strengths in both
linear absorption and emission spectra of polaritons have already been analyzed
theoretically~\cite{delPino2015,neuman2018}, although in~\cite{neuman2018} using
exclusively Lindblad operators. Instead, our approach using Markovian BRW theory
introduces the asymmetry more naturally through the function $S_B(\omega)$ and
provides a more consistent approach as implemented in~\cite{delPino2015} for
absorption. In \autoref{fig:Figure6} we include the linear absorption and
excitation-emission spectra for a TC model with $N=2$ emitters. When the
dephasing process is implemented as a Lindblad operator, both spectra are highly
symmetrical. With the BRW theory the upper polariton absorbs less than the lower
one and the emission spectra display a strong asymmetry where the U/L peak
dominates. This reflects the importance of correctly representing the decay
mechanisms of polaritons. In these linear spectroscopies, the states involved
are the ground state $G$ and those of the first excitation manifold $\{L,D,U\}$.
The upper polariton $U$ decays to the ground state $G$ by photon loss but also
to the lower polariton $L$ and the dark states due to vibrational relaxation, whereas the lower polariton decays only to the ground state (at low
enough temperature or high enough Rabi splitting). Thus the upper polariton peak
is broadened in the absorption, and in the emission spectrum the path of laser
excitation of the upper polariton followed by relaxation to the lower polariton
produces the dominant U/L peak. However, these linear spectroscopies do not
provide a detailed understanding on the inner dynamics of the polariton states
involved and below we study these emerging asymmetries beyond the linear
response.

\subsection{Non-linear 2D spectroscopy}

\begin{figure}[t]
\includegraphics[width=\linewidth]{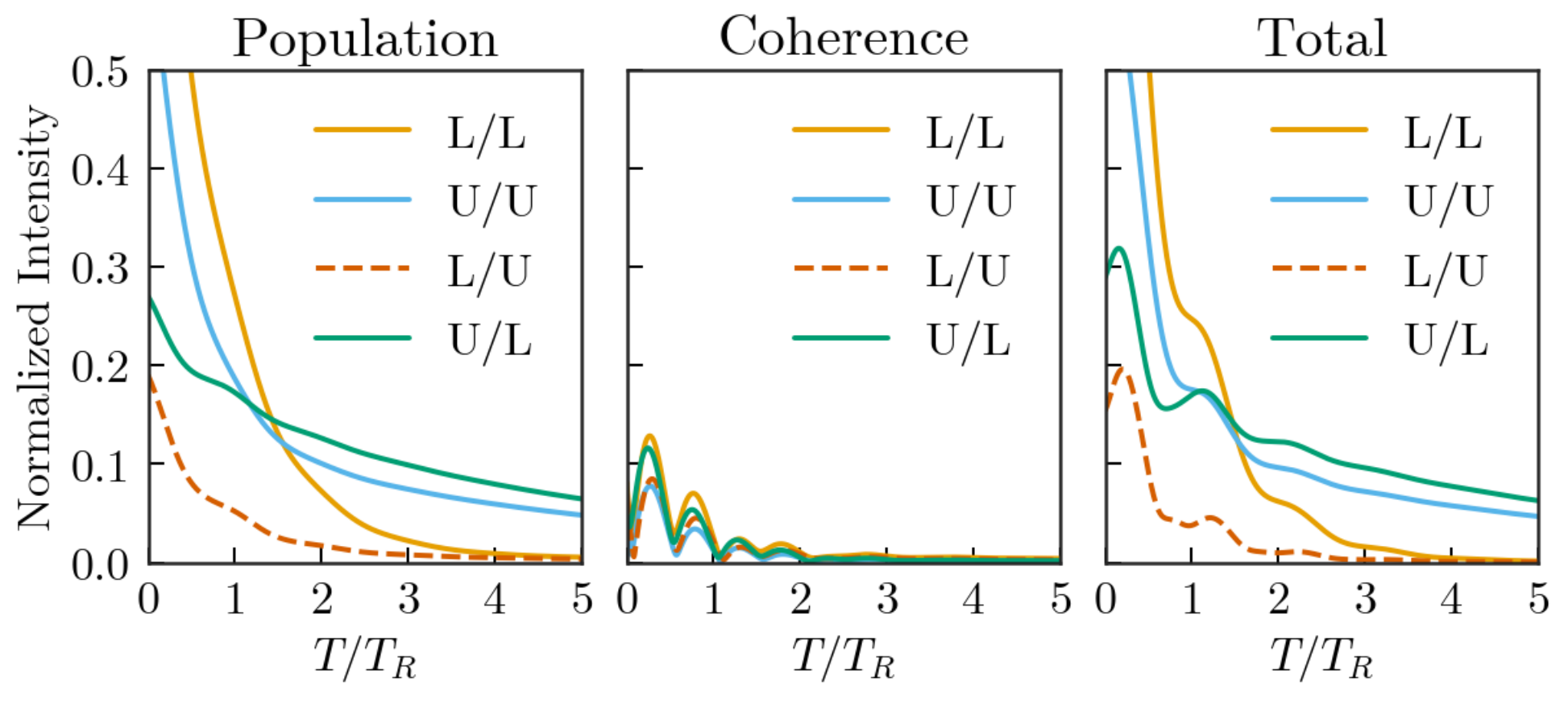}
\caption{ \label{fig:Figure7}
Time evolution of the diagonal peaks \{L/L,U/U\} and cross peaks \{L/U, U/L\} in the two dimensional spectrum $|S(\omega_t,T,\omega_\tau)|$ for a system of $N=2$ molecules
in resonance with the cavity $\omega_0=\omega_c=2$ eV, Rabi frequency $\Omega_R=0.1$ eV, cavity lifetime 15 fs and dephasing lifetime 50 fs.
The chosen simple spectral density function here is $J(\omega)=\gamma$ at zero temperature.
The components corresponding to populations (exponential) and coherences (oscillatory) are separated in the two first panels.
}
\end{figure}

In \autoref{sec:Formula2DS} we derived a formula for the 2D spectral function
$S(\omega_t, T, \omega_\tau)$ by which the 2DS at each pair of frequencies
$(\omega_\tau , \omega_t)$ can be understood as built from the interference of
damped oscillations, $e^{\lambda_j T} = e^{-\Gamma_j T} e^{i \omega_j T}$, with
different amplitudes $E_j(\omega_{\tau}) D_j(\omega_t)$. A detailed analysis
reveals that only for a few Liouvillian eigenstates neither the excitation nor
the detection mask functions vanish, and only these contribute to the sum over
$j$ in \autoref{eq:spectra}. In our polariton showcase, these states are related to Hamiltonian populations
$\{LL, UU, DD\}$ and coherences $\{LU, UL\}$. The temporal factor $e^{-\Gamma_j
T}$ causes a leading exponential decay during the waiting time $T$, while the
factor $e^{i \omega_j T}$ explains the ubiquitous presence of Rabi oscillations
in the 2DS, since the only contributing frequencies $\omega_j \sim \pm
\omega_{UL}$ correspond to coherences that involve the lower $L$ and upper $U$
polaritons (see \autoref{fig:Figure7} where we illustrate the crucial importance
of coherences during the waiting time $T$ since they provide most of the
oscillatory components). As analyzed in the theory
\autoref{sec:Liouvillecomplex}, the location of the eigenvalues in the complex
plane helps to understand the relative global decay of populations, showing that
the upper polariton decays faster because it has three available decay channels,
($U \to G$ due to photon loss and $U \to D$ and $U \to L$ due to vibrational
bath relaxation). Both the lower polariton and the dark states have a single
decay channel each ($L \to G$ due to photon loss and $D \to L$ due to
vibrational relaxation). In many experimental implementations with thin metallic
mirrors, photon loss lifetimes are shorter than molecular dephasing
lifetimes~\cite{fregoni2022}, and we here choose 15 fs for $1/\kappa$ and 50 fs
for $1/\gamma$. However, note that for the comparison of our results with the
experimental ones of~\cite{mewes2020} in \autoref{sec:experiment}, the reverse
criterion ($1/\kappa > 1/\gamma$) is required for improved agreement.

\begin{figure}[t]
\centering
\includegraphics[width=\linewidth]{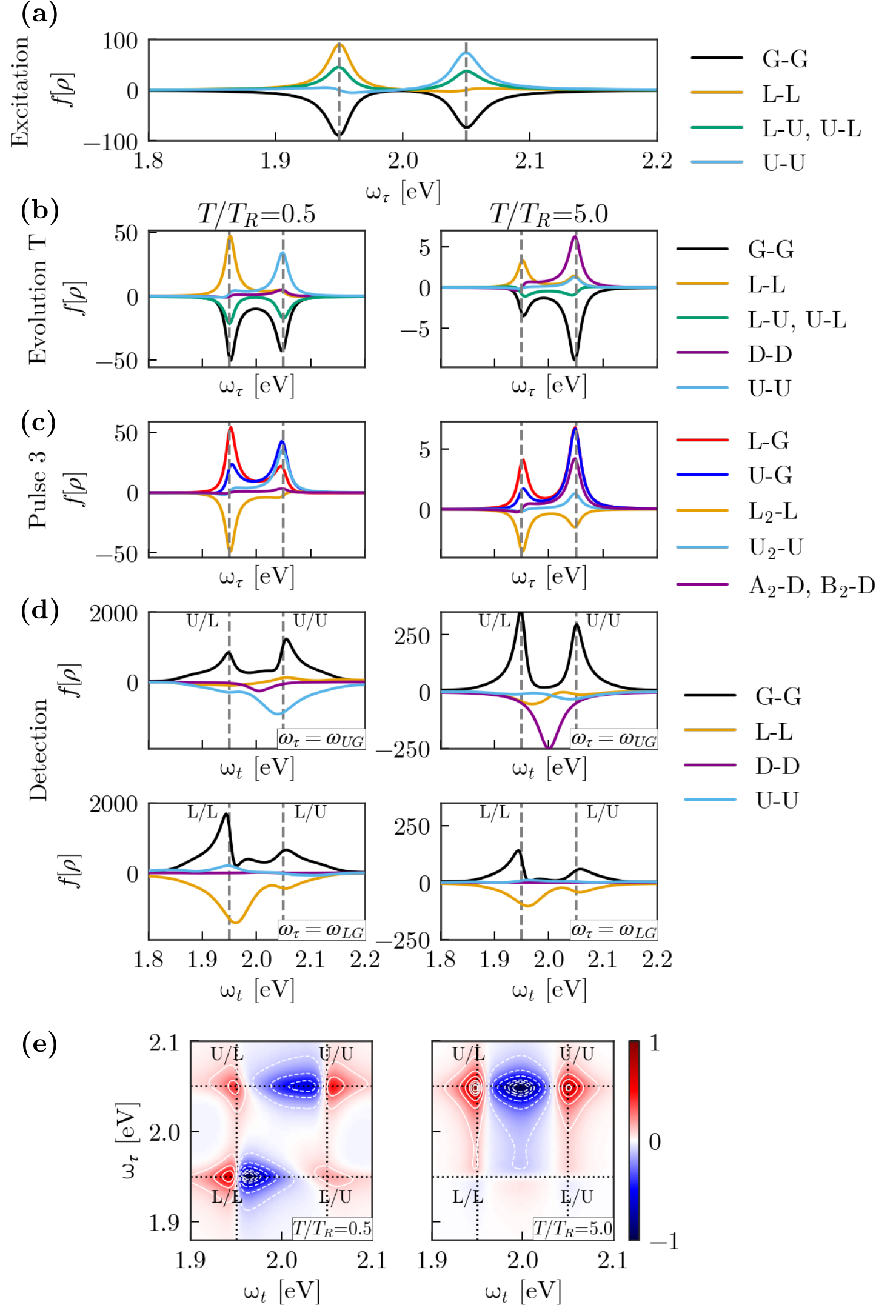}
\caption{ \label{fig:Figure8}
Temporal build up of the 2DS for a TC model with $N = 2$ in a vertical time-line with four stages: (a) excitation [pulse 1, $\tau$, pulse 2], (b) waiting $T$, (c) pulse 3 and (d) detection.
In the figures $f(\rho)$ indicates the Fourier transform of Re$[\rho (t)]$ for the rephasing plus non-rephasing contributions. Solid lines indicate the non-zero components of
$f(\rho)$ in terms of the Hamiltonian eigenstates in the three excitation manifolds $\Lambda_0, \Lambda_1$ and $\Lambda_2$. Vertical dashed lines indicate the positions of the transition frequencies
$\omega_{LG}=1.95$ eV and $\omega_{UG}=2.05$ eV. Two waiting times are considered, $T/T_R=0.5$ and $T/T_R=5.0$. At the detection stage, $f(\rho)$ is plotted against $\omega_t$ and
the solid line of the component G-G corresponds exactly to the 2DS along the cuts $\omega_{\tau}=\omega_{LG}$ and $\omega_{\tau}=\omega_{UG}$.
(e) 2DS absorptive spectrum $S_{\text{Abs}}(\omega_t,T,\omega_\tau)$ for the two waiting times.
The system and parameters for this calculation are the same as described in \autoref{fig:Figure7}. Spectra are normalized to unity at their maxima.
}
\end{figure}

For the open TC systems, a major feature is found in the computed 2DS as the
waiting time $T$ increases (see \autoref{fig:Figure7}); they develop strong
asymmetries during the waiting time $T$, revealed by the relative intensities of
cross peaks ($L/U < U/L$), with the diagonal peak $L/L$ displaying the fastest
decay (see \autoref{fig:Figure7}). A similar behavior is found experimentally
for J-aggregates within optical cavities~\cite{mewes2020}, as discussed below.
In contrast, for a non-dissipative TC model the 2DS shows identical diagonal
peak intensities (L/L and U/U) in temporal counterphase with identical
cross-peaks heights (U/L and L/U) for all waiting times $T$ and the 2DS show
full revivals when $T/T_R$ is integer (Rabi oscillations).

To understand the origin of these asymmetries in 2DS, in \autoref{fig:Figure8}
we analyze the calculation step by step for a dissipative TC model with $N=2$.
Intermediate temporal expressions for the density $\kket*{\rho}$ help us to
track the history at each step. Note that the perturbative density matrix
$\rho(t)$ is not Hermitian and is not represented by a positive definite
matrix, since $\rho^{(3)}$ for a given phase matching condition involves
commutators with non-Hermitian coupling interactions $\hat{\mu}^+$ and
$\hat{\mu}^-$.

\begin{figure*}[t]
\includegraphics[width=\linewidth]{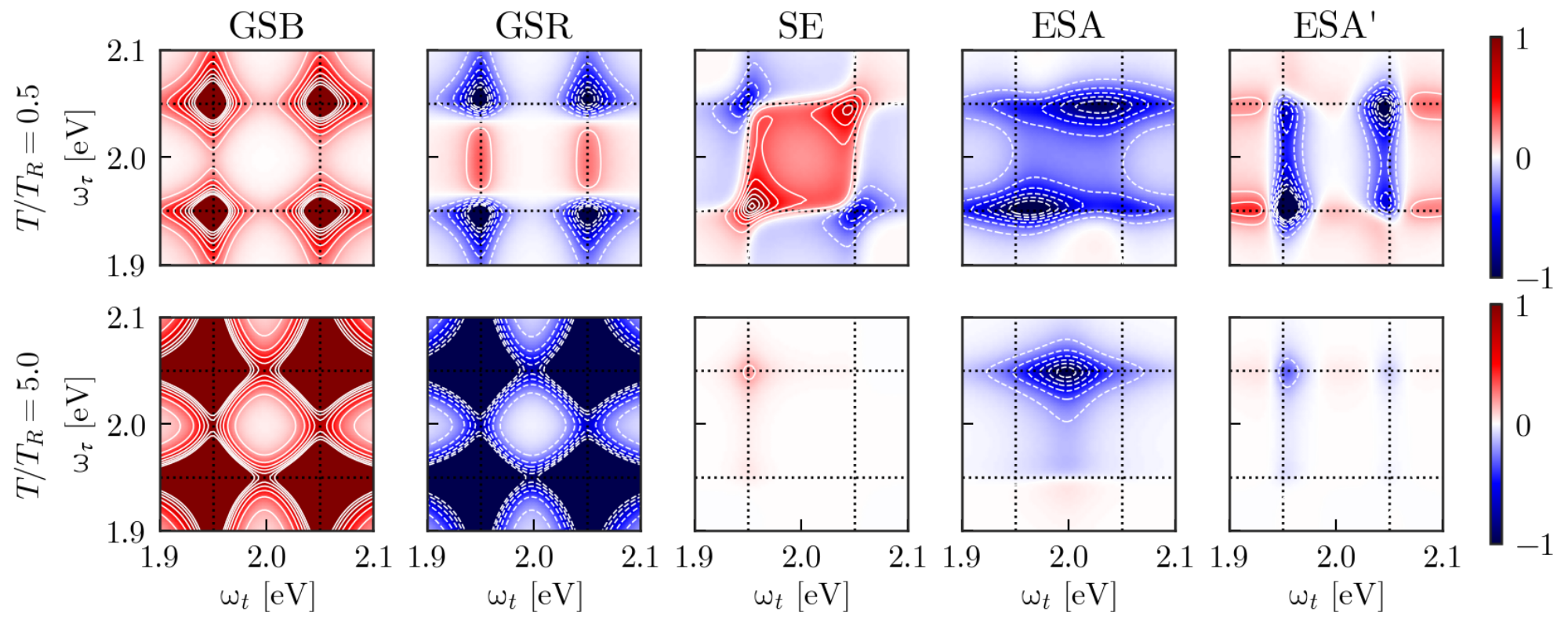}
\caption{ \label{fig:Figure9}
Components GSB, GSR, SE, ESA and ESA$^\prime$
of the total absorptive spectrum $S_{\text{Abs}} (\omega_t, T, \omega_\tau)$ plotted in \autoref{fig:Figure8}(e) for two different waiting times $T/T_R=0.5$ and
$T/T_R=5.0$, related to the Feynman paths quoted in \autoref{fig:Figure4}(b). Total spectra are normalized to unity at their maxima.
It is worth noting that 2D plots always contain both population and coherence components, so that GSR and SE shows both positive and negative signals at short $T$. At large $T$ GSR is negative and it compensates the positive GSB and the residual SE becomes positive.
}
\end{figure*}

\begin{enumerate}
\item The first pulse takes the ground state population $GG$ to the coherences
$\{GL,GU\}$ [rephasing case, see \autoref{fig:Figure4}(b)] and $\{LG,UG\}$
(non-rephasing). These coherences form independent diagonal blocks in the
Liovillian representation and do not mix.

\item Time evolution $\tau$: The coherences evolve within their matrix blocks
with a phase $e^{\lambda_j \tau}$, where $\lambda_j$ contains frequencies
$\omega_j \sim \pm \omega_{GU}$ or $\omega_j \sim \pm \omega_{GL}$, so that
signals in the 2DS are expected at both frequencies. From the Liouvillian master
equations we extract $\dot{\rho}_{LG} = -(\kappa/2) \rho_{LG} + (\gamma/8-\kappa/2) \rho_{UG}$ and
$\dot{\rho}_{UG} = -(\kappa/2) \rho_{LG} + (-\gamma/8-\kappa/2) \rho_{UG}$. Thus the dephasing (with the parameter $\gamma$) makes the coherence $UG$ decay slightly
faster than the $LG$ one.

\item The second pulse takes the previous coherences to populations
$\{GG,LL,UU\}$ and coherences $\{LU,UL\}$ (contained in another diagonal block
in the Liouvillian matrix). The Fourier transform of $\Re[\rho(t)]$ up to this
step is plotted in \autoref{fig:Figure8}(a). Whereas the population $LL$
inherits a combination $\propto -(LG-GL)$ from step 2 (and consequently produces
a signal at frequency $\omega_\tau=\omega_{LG}$ but zero at $\omega_{UG}$), the
population $UU$ comes from $\propto - (UG+GU)$ (with a peak at frequency
$\omega_\tau=\omega_{UG}$ but zero at $\omega_{LG}$). Therefore both signals
must be asymmetric with respect to frequencies $\omega_{LG}$ and $\omega_{UG}$.
The GG population results from the combination $\propto + (UG-LG+GL-GU)$ that
carries both frequencies (almost yields a symmetric signal) and has opposite
sign with respect to the other populations. A similar reasoning follows for the
coherences $LU$ and $UL$. Since the coherence $UG$ decays faster during the
evolution with $\tau$, the peaks at $\omega_\tau = \omega_{UG}$ show a lower
intensity here.

 \item Evolution during $T$. In \autoref{fig:Figure8}(b), we show two different
 waiting times. For small times, $T/T_R=0.5$, the appearance of the dark state
 population $DD$ and the sign inversion of the coherences $\{LU,UL\}$ can
 already be noticed. At longer times, $T/T_R=5.0$, the populations decay at
 $\omega_{LG}$, but those of the ground and dark states prevail at
 $\omega_{UG}$. Here it can be appreciated that the ground state population
 increases roughly with the exponential decay of lower polariton population
 $LL$, at $\omega_{LG}$. In contrast, at $\omega_{UG}$ the ground state
 population increases due to the upper polariton $U$ decay, but to a lesser
 extend also because $U$ decays into the dark state $D$. This is in fact the
 crucial step due to which the $L$ vs.\ $U$ asymmetry eventually appears in the
 2DS\@.

 \item Pulse 3 causes the transfer to the coherence blocks $\{ LG,UG \}$ and $\{
 \Lambda_2\Lambda_1 \}$. From the previous step, $\{ GG,LL,LU \}$ contribute to
 the coherence $LG$ and $\{GG,UL,UU\}$ do so for the coherence $UG$. This
 explains the components seen in \autoref{fig:Figure3}(c). Specifically for long
 waiting times the shapes of $\{ LG,UG \}$ coherences originate from the
 dominant ground state in \autoref{fig:Figure8}(b). Also, population $DD$ is
 excited to coherences $\Lambda_2 D$. Similarly, $LL$ transfers to $L_2L$ and
 $UU$ to $U_2 U$.

\item Signal is detected during the evolution in time $t$: the coherences $\{
 LG,UG \}$ decay to the ground state population $GG$ and coherences
 $\{\Lambda_2L, \Lambda_2U, \Lambda_2 D \}$ decay radiatively to the $\Lambda_1$
 populations $LL,DD$ and $UU$, plotted in \autoref{fig:Figure8}(d) along the
 excitation cuts $\omega_{\tau} = \omega_{UG}$ and $\omega_{\tau} =
 \omega_{LG}$. The sum of these populations gives the structures observed in the
 2DS in \autoref{fig:Figure8}(e).
 \end{enumerate}

When exciting at the upper polariton frequency $\omega_\tau=\omega_{UG}$, the
most important negative contribution to the 2DS comes from the population $UU$ at short
waiting times, whereas at long waiting times the $U$ state decays into the dark
states and the $DD$ population dominates. In contrast, for lower
polariton excitation $\omega_\tau=\omega_{LG}$, the dynamics of the negative contribution to the 2DS is governed
by the $LL$ population at any waiting time, because dark states hardly
contribute to the dynamics and all populations have similar shapes during $T$
(only the intensity is reduced exponentially as expected). Whereas the
population contributions tend to approximately cancel to each other for lower
polariton excitation as $T$ increases, this does not occur when pumping the
upper polariton, which has a different mechanism that is contributed by the dark
state. Ultimately, this explains why at long waiting times the U/L and U/U peaks
remain visible against the more rapidly vanishing L/L and L/U peaks in the
2DS\@.

To understand these asymmetries in another way, we show in \autoref{fig:Figure9}
the decomposition of the total 2DS of \autoref{fig:Figure8}(e) into all Feynman
path components GSB, GSR, SE, ESA and ESA$^\prime$ for the same two waiting
times. The GSB positive contribution remains constant during the waiting time,
while GSR grows (with opposite sign) since any state finally decays into the
ground state. For long waiting times the $LL$ population fully relaxes by photon
loss after excitation into the ground state and GSR roughly produces the same
signal as the GSB component but with opposite sign, thus producing a complete
cancellation in the spectra at $\omega_\tau=\omega_{LG}$ (lower peaks L/L and
L/U). At variance, the $UU$ population may decay by dephasing relaxation to the
dark D state and lower L state, and to the ground state by photon loss. Thus the
residual $GG$ population to be excited by the third pulse in GSR becomes smaller
than in GSB, which leads to only a partial cancellation in GSB+GSR at the upper
peaks U/L and U/U. In the negative ESA component, the third pulse makes the
difference when exciting at $\omega_\tau=\omega_{LG}$ or
$\omega_\tau=\omega_{UG}$. In the former case, the coherence $L_2L$ moves back
to the $LL$ population in detection, aided only by a cavity photon loss; in the
latter case the coherence $U_2U$ decays into $UU$ and subsequently into $DD$
populations at detection, aided by both molecular relaxation and cavity photon
loss. Eventually, the $DD$ population produces the remnant ESA signal, already
present for $T/T_R=5.0$.

The contribution of the five different components (GSB, SE, ESA, GSR and
ESA$^\prime$) to the 2DS depends upon the waiting time $T$. In this respect both
components SE and ESA$^\prime$ are crucial in the construction of the full 2DS
but they tend to vanish at large waiting times due to the relaxation of
populations and coherences from the $\Lambda_1$ and $\Lambda_2$ excitation
manifolds. The mechanism of the ESA$^\prime$ component corresponds to relaxation
between coherences of different excitation manifolds from $\ketbra*{\Lambda_2}{\Lambda_1}$ to $\ketbra*{\Lambda_1}{G}$. In \autoref{fig:Figure9} both
contributions ESA and ESA$^\prime$ clearly contribute to the negative part of
the spectrum at short waiting time, $T/T_R=0.5$, while SE and ESA$^\prime$ components
tend to vanish for the positive and negative part of the spectra, respectively,
for $T/T_R=5$ and ESA remains the main negative contribution.

\begin{figure}[t]
\includegraphics[width=\linewidth]{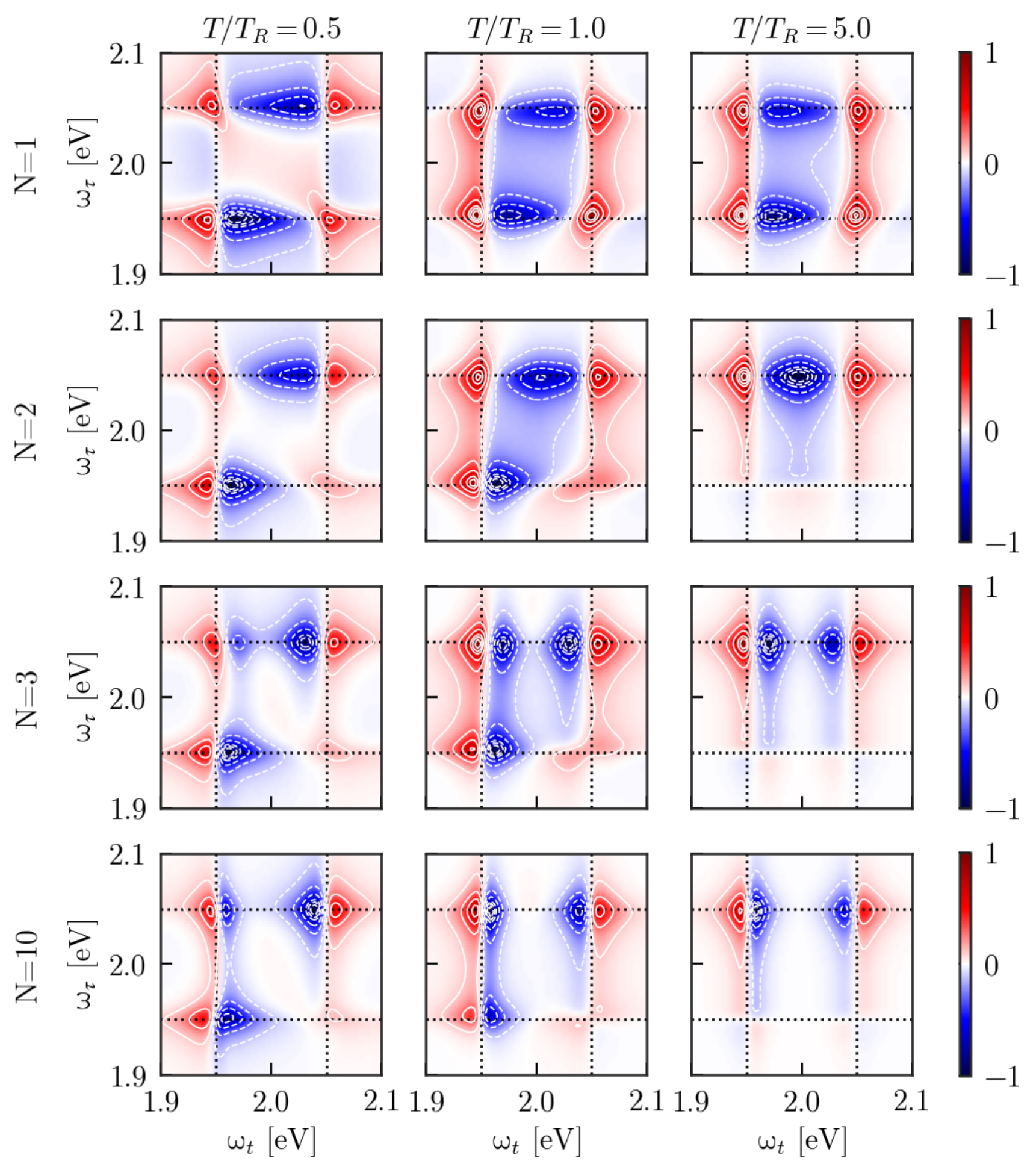}
\caption{
\label{fig:Figure10}
2DS for the dissipative Tavis-Cummings system for a different number of molecules $N =1, 2, 3$ and 10, and for three different waiting times $T/T_R$ = 0.5, 1.0 and 5.0.
The parameters are the same as quoted in \autoref{fig:Figure9}. All 2DS are normalized to unity at their maxima.
}
\end{figure}

In addition to the case with $N=2$ molecules discussed above, we have also
studied the 2DS for $N=1$ (Jaynes-Cummings) and for $N=3$ and $N=10$. Concerning
the case $N=1$ we do not observe the same fading trends at large $T$ in the 2DS
(see \autoref{fig:Figure10}). In the absence of a dark state, here the upper
polariton still has two decay mechanisms (radiative $U \to G$ and vibrational $U
\to L$) so that at long $T$ the GSR signal for excitation at
$\omega_{\tau}=\omega_{UG}$ should be larger than for excitation at
$\omega_{\tau}=\omega_{LG}$. However, this difference is very small compared
with the case $N>1$ with dark states. As $T$ increases, the GSR signals at any
of the peaks fully cancel the constant GSB, and this happens almost
simultaneously. Additionally, a shift of the peak energy of the negative ESA
components is visible along the excitation cut $\omega_{\tau}=\omega_{UG}$
during $T$. At short time $T$, the ESA peak is centered at the $U_2 \to U$
transition. However, due to the dissipative $U \to L$ decay, for large $T$ the
third pulse mainly excites the $L$ state to the second excitation manifold, with
the subsequent ESA peak centered at the transition $L_2 \to L$.

For the case of $N>2$, as $N$ increases, the energies of states $\{A_2\}$ and
$\{B_2\}$ in $\Lambda_2$ get closer to those of the $L_2$ and $U_2$ polaritons,
respectively. Thus the ESA contributions for $N > 2$ produce negative signals
approaching the detection frequencies $\omega_{\tau}=\omega_{LG}$ and
$\omega_{\tau}=\omega_{UG}$ (see \autoref{fig:Figure10}). Note also that these
states in $\Lambda_2$ are radiatively connected to the $N-1$ dark states in
$\Lambda_1$. However, the mechanism involving the collection of dark states
remains unaltered and the conclusions reached for $N=2$ are also valid for large
$N$. Our results indicate that the shape of 2DS remains unchanged for $N \ge 3$,
while the global intensity reduces with increasing $N$. In the thermodynamic
limit $N\to\infty$, the TC model becomes linear and can be represented by two
coupled harmonic oscillators~\cite{garraway2011}, such that the anharmonicity
that leads to a nonzero 2DS disappears. However, a realistic treatment of 2DS
for large systems would require going beyond the third-order perturbative limit,
as the number of absorbed photons would also increase with $N$ for a given
driving strength~\cite{mukamel1999}, reaching much higher excitation manifolds
$\Lambda_n$, with the nonlinearity scaling as $n / N$. The system is then
expected to behave similarly as a TC model with a small number of molecules, as
implied by the reasonable agreement with experiments (see \autoref{fig:Figure11}
below).

\subsection{Comparison with experiments}

\label{sec:experiment}
\begin{figure}[tbp]
\includegraphics[width=0.93\linewidth]{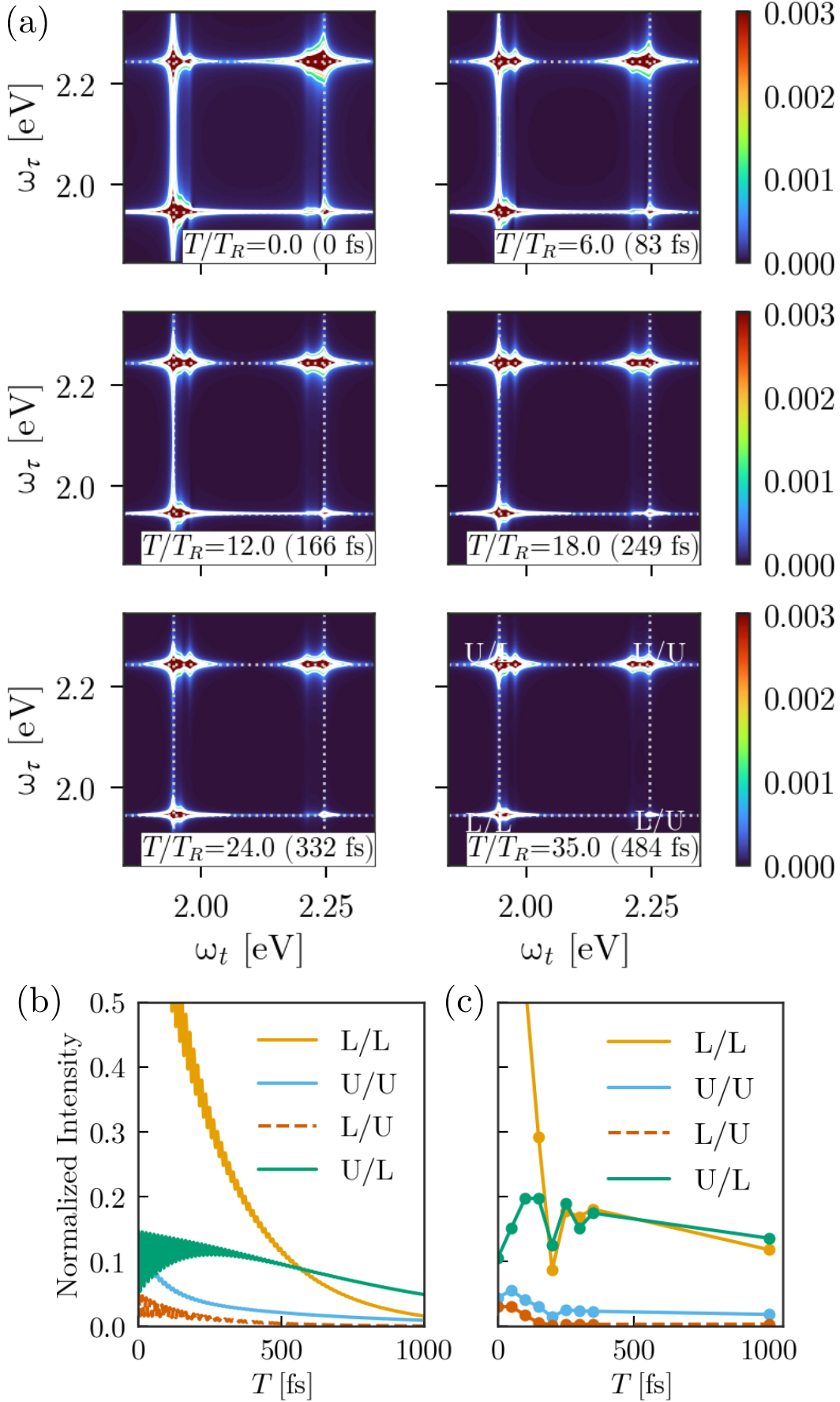}
\caption{ \label{fig:Figure11}
(a) 2DS $| S(\omega_t, T , \omega_\tau)|$ for a TC model with $N=5$ emitters, for different waiting times $T$
(in units of the Rabi period $T_R=1/\Omega_R=13.8$ fs). The 2DS at waiting time $T/T_R=0$ is normalized to unity at its maximum value and those for $T/T_R >0$ share the same scale as in $T/T_R=0$.
The molecular system has a small positive detuning with $\omega_c=2.1$ eV and $\omega_0=2.09$ eV, and a Rabi splitting $\Omega_R=0.3$ eV.
Dotted lines indicate the positions for the transition frequencies $\omega_{LG}=1.945$ eV and $\omega_{UG}=2.245$ eV.
Cavity lifetime is 120 fs and molecular dephasing lifetime is chosen 60 fs and we have also considered a Debye spectral function (with cut-off $\delta=0.2$ eV) at room temperature ($T=300$ K).
(b) Time evolution of the diagonal and cross peaks maxima in (a) in terms of the waiting time $T$. The peak heights are renormalized such that the value of the highest peak $L/L$ at $T=0$ is unity.
(c) Experimental data extracted from diagonal and cross peaks in the 2DS in~\cite{mewes2020} corresponding to a system of J-aggregates
within an optical cavity.
}
\end{figure}

Recent experiments with a polaritonic system involving molecular
J-aggregates~\cite{mewes2020} inside a microcavity show three main features in
2DS as a function of increasing waiting time $T$ (see \autoref{fig:Figure11}),
namely,
\begin{itemize}
\item the 2DS rapidly develops an asymmetry where the U/L cross peak gains in
intensity, while the L/U cross peak disappears.
\item the diagonal peak L/L, initially the most intense, has the fastest decay
and drops to an intensity level comparable to the U/L cross peak.
\item the peaks L/L and U/L dominate compared to the U/U and L/U peaks.
\end{itemize}

To reproduce these features, we study a dissipative TC model with $N=5$ molecules (larger
ensembles do not show distinguishable differences in the 2DS, as discussed
above) with a two-level natural frequency $\omega_e=2.09$~eV and with a slightly
blue-shifted cavity mode, $\omega_c=2.1$~eV. As in the experiment, the Rabi
frequency is chosen as $\Omega_R=0.3$~eV. To reproduce the experimental
conditions more realistically, temperature effects are also incorporated in the
exciton-phonon coupling through a Debye spectral function for the bath at room
temperature $T=300$~K and using a frequency cut-off parameter $\delta=0.2$ eV.
We also find that the cavity lifetime (120 fs) must be set somewhat longer than
the dephasing lifetime (60 fs) to reproduce the three experimental features
listed above. The source of these phenomena stems from the relaxation mechanisms
of states in $\Lambda_1$ and $\Lambda_2$ manifolds as explained above. Of
course, the comparison of the simple model treated here with the experiment is
not perfect. J-aggregates have interactions with the environment and between
molecules in addition to a phonon structure that might cause a slower decay of
peaks than that produced by the electronic motion alone. Oscillations in the L/L
and U/L peaks vs $T$ present in the experiment below 500 fs (\autoref{fig:Figure11}(c)) cannot be attributed
to electronic Rabi oscillations (which in fact are much faster and visible in
the theoretical peaks L/U and U/L below 500 fs in \autoref{fig:Figure11}(b)). 
We speculate that a more detailed account of the vibrational structure and its influence on the polariton modes (in the current model determined by the single parameter $\gamma$) would be required to reproduce these experimental features while keeping a shorter cavity lifetime. However, despite the simplicity of our model, the reproduction of the trends of these three experimental
features is remarkable.

It would be useful to extract from the experimental 2DS some physical parameters, like partial decay rates for the processes involved in the polariton photodynamics. The time-evolution of diagonal or cross peaks
against the waiting time already contains information on the decays. From our expression for 2DS \autoref{eq:spectra} we can extract the value of the spectra at any frequency point
$(\omega_\tau, \omega_t)$. For the L/L diagonal peak the analytical expression for the peak intensity reads
\begin{align}
& | S(\omega_L, T, \omega_L)|^2 = C_1 e^{- 2 \Gamma_{LL} T} + e^{- (\Gamma_{LL}+\Gamma_{UL}) T} \\ \nonumber
& \times [C_2 \cos(\Omega_R T) + C_3 \sin(\Omega_R T)] + e^{- 2 \Gamma_{UL} T} \\ \nonumber
& \times \left[ C_4 \cos^2 (\Omega_R T) + C_5 \sin^2 (\Omega_R T) + C_6 \sin(2 \Omega_R T) \right]
\end{align}
where $\Gamma_{LL}= \kappa/2$ and $\Gamma_{UL}=\kappa/2 + \gamma/8$. From the latter equation we have worked out a simplified expression to fit the 2DS L/L peak.
\begin{align}
 & | S(\omega_L, T, \omega_L) |^2 \sim \nonumber \\
 & | A e^{-2 \Gamma_{LL} T} + e^{- 2 \Gamma_{UL} T} [B \cos(\Omega_R T) + C \sin(\Omega_R T)] |.
 \end{align}
For instance, for the 2DS obtained for the JC model in
\autoref{fig:Figure10}, using $\kappa=0.044$ eV (lifetime 15 fs) for cavity
photon loss and $\gamma=0.0132$ (lifetime 50 fs) for dephasing, the fitting of
the L/L peak against $T$ with this 5 parameter function $(A,B,C,\kappa, \gamma)$
(Rabi frequency $\Omega_R$ is extracted directly from the oscillations in the
plot) yields 15 and 43 fs respectively.

\section{Conclusions}\label{sec:Conclusions}

In this work, we have analyzed the structure and dynamics of cavity polaritons
derived from fundamental models like Jaynes-Cummings and Tavis-Cummings
Hamiltonians combined with a perturbative treatment of exciton-vibration
interactions in order to understand the outcome of multidimensional coherent
spectroscopy when applied to molecules. We have derived an efficient
pseudo-analytic procedure to compute 2DS for three non-overlapping laser pulses
that allows to interpret the two-dimensional spectrum by analyzing their
build-up at any stage, namely, excitation, evolution, probe and detection.

Asymmetries in experimental 2DS concerning diagonal and cross peaks are here
explained by the crucial role played by dark states within the first excitation
manifold. We show the relevance of components GSR and ESA$^\prime$ that involve
relaxation in addition to the standard GSB, SE, ESA components. The dynamic role
of coherences is to bring oscillatory patterns to the spectra, on top of the
background contributed by populations, and is clearly relevant during the
waiting time between pump and probe. Finally, some emerging features in the ESA
path produce a directly detected signal in the 2DS that conforms a fingerprint
of dark states.

The 2D experimental spectra with J-aggregates has more complex structures and
features that cannot be reproduced fairly with our simplified model. Further
developments require to include molecular vibrational states with their
corresponding anharmonicity, disorder and broadening effects that eventually
contribute at longer waiting times beyond the fast electronic photodynamics.
Also we think our method and conclusions may find applications in other
scenarios, for instance, in semiconductor microcavities~\cite{takemura2015} or
to understand relaxation in plexcitonic materials~\cite{finkelstein2021}. We
hope this work may contribute to the design of new experiments with electronic
polaritons involving molecular ensembles in microcavities, with emphasis on
preparing the active system with just a few atoms or molecules, which by itself
is quite an experimental challenge.

\begin{acknowledgments}
We acknowledge funding by the Vicerrectoría de Investigación at Universidad de
Antioquia under project CODI Programática 2022-53576, by the Spanish Ministry
for Science and Innovation-Agencia Estatal de Investigación (AEI) through Grants
PID2021-125894NB-I00, CEX2018-000805-M (through the María de Maeztu program for Units of Excellence in R\&D), and by the European Research
Council through Grant No. ERC-2016-StG-714870. L. M. acknowledges the Swiss
National Science Foundation, project P2ELP2\_187957.
\end{acknowledgments}

\bibliography{Polaritons}

\providecommand{\noopsort}[1]{}\providecommand{\singleletter}[1]{#1}%
\begin{thebibliography}{46}%
\makeatletter
\providecommand \@ifxundefined [1]{%
 \@ifx{#1\undefined}
}%
\providecommand \@ifnum [1]{%
 \ifnum #1\expandafter \@firstoftwo
 \else \expandafter \@secondoftwo
 \fi
}%
\providecommand \@ifx [1]{%
 \ifx #1\expandafter \@firstoftwo
 \else \expandafter \@secondoftwo
 \fi
}%
\providecommand \natexlab [1]{#1}%
\providecommand \enquote  [1]{``#1''}%
\providecommand \bibnamefont  [1]{#1}%
\providecommand \bibfnamefont [1]{#1}%
\providecommand \citenamefont [1]{#1}%
\providecommand \href@noop [0]{\@secondoftwo}%
\providecommand \href [0]{\begingroup \@sanitize@url \@href}%
\providecommand \@href[1]{\@@startlink{#1}\@@href}%
\providecommand \@@href[1]{\endgroup#1\@@endlink}%
\providecommand \@sanitize@url [0]{\catcode `\\12\catcode `\$12\catcode `\&12\catcode `\#12\catcode `\^12\catcode `\_12\catcode `\%12\relax}%
\providecommand \@@startlink[1]{}%
\providecommand \@@endlink[0]{}%
\providecommand \url  [0]{\begingroup\@sanitize@url \@url }%
\providecommand \@url [1]{\endgroup\@href {#1}{\urlprefix }}%
\providecommand \urlprefix  [0]{URL }%
\providecommand \Eprint [0]{\href }%
\providecommand \doibase [0]{https://doi.org/}%
\providecommand \selectlanguage [0]{\@gobble}%
\providecommand \bibinfo  [0]{\@secondoftwo}%
\providecommand \bibfield  [0]{\@secondoftwo}%
\providecommand \translation [1]{[#1]}%
\providecommand \BibitemOpen [0]{}%
\providecommand \bibitemStop [0]{}%
\providecommand \bibitemNoStop [0]{.\EOS\space}%
\providecommand \EOS [0]{\spacefactor3000\relax}%
\providecommand \BibitemShut  [1]{\csname bibitem#1\endcsname}%
\let\auto@bib@innerbib\@empty
\bibitem [{\citenamefont {Herrera}\ and\ \citenamefont {Spano}(2016)}]{herrera2016}%
  \BibitemOpen
  \bibfield  {author} {\bibinfo {author} {\bibfnamefont {F.}~\bibnamefont {Herrera}}\ and\ \bibinfo {author} {\bibfnamefont {F.~C.}\ \bibnamefont {Spano}},\ }\bibfield  {title} {\bibinfo {title} {Cavity-controlled chemistry in molecular ensembles},\ }\href {https://doi.org/10.1103/PhysRevLett.116.238301} {\bibfield  {journal} {\bibinfo  {journal} {Physical Review Letters}\ }\textbf {\bibinfo {volume} {116}},\ \bibinfo {pages} {238301} (\bibinfo {year} {2016})}\BibitemShut {NoStop}%
\bibitem [{\citenamefont {Ribeiro}\ \emph {et~al.}(2018)\citenamefont {Ribeiro}, \citenamefont {Mart{\'\i}nez-Mart{\'\i}nez}, \citenamefont {Du}, \citenamefont {Campos-Gonzalez-Angulo},\ and\ \citenamefont {Yuen-Zhou}}]{ribeiro2018}%
  \BibitemOpen
  \bibfield  {author} {\bibinfo {author} {\bibfnamefont {R.~F.}\ \bibnamefont {Ribeiro}}, \bibinfo {author} {\bibfnamefont {L.~A.}\ \bibnamefont {Mart{\'\i}nez-Mart{\'\i}nez}}, \bibinfo {author} {\bibfnamefont {M.}~\bibnamefont {Du}}, \bibinfo {author} {\bibfnamefont {J.}~\bibnamefont {Campos-Gonzalez-Angulo}},\ and\ \bibinfo {author} {\bibfnamefont {J.}~\bibnamefont {Yuen-Zhou}},\ }\bibfield  {title} {\bibinfo {title} {Polariton chemistry: controlling molecular dynamics with optical cavities},\ }\href {https://doi.org/https://doi.org/10.1039/C8SC01043A} {\bibfield  {journal} {\bibinfo  {journal} {Chemical science}\ }\textbf {\bibinfo {volume} {9}},\ \bibinfo {pages} {6325} (\bibinfo {year} {2018})}\BibitemShut {NoStop}%
\bibitem [{\citenamefont {Feist}\ \emph {et~al.}(2018)\citenamefont {Feist}, \citenamefont {Galego},\ and\ \citenamefont {Garcia-Vidal}}]{feist2018}%
  \BibitemOpen
  \bibfield  {author} {\bibinfo {author} {\bibfnamefont {J.}~\bibnamefont {Feist}}, \bibinfo {author} {\bibfnamefont {J.}~\bibnamefont {Galego}},\ and\ \bibinfo {author} {\bibfnamefont {F.~J.}\ \bibnamefont {Garcia-Vidal}},\ }\bibfield  {title} {\bibinfo {title} {Polaritonic chemistry with organic molecules},\ }\href {https://doi.org/https://doi.org/10.1021/acsphotonics.7b00680} {\bibfield  {journal} {\bibinfo  {journal} {ACS Photonics}\ }\textbf {\bibinfo {volume} {5}},\ \bibinfo {pages} {205} (\bibinfo {year} {2018})}\BibitemShut {NoStop}%
\bibitem [{\citenamefont {Fregoni}\ \emph {et~al.}(2022)\citenamefont {Fregoni}, \citenamefont {Garcia-Vidal},\ and\ \citenamefont {Feist}}]{fregoni2022}%
  \BibitemOpen
  \bibfield  {author} {\bibinfo {author} {\bibfnamefont {J.}~\bibnamefont {Fregoni}}, \bibinfo {author} {\bibfnamefont {F.~J.}\ \bibnamefont {Garcia-Vidal}},\ and\ \bibinfo {author} {\bibfnamefont {J.}~\bibnamefont {Feist}},\ }\bibfield  {title} {\bibinfo {title} {Theoretical challenges in polaritonic chemistry},\ }\href {https://doi.org/https://doi.org/10.1021/acsphotonics.1c01749} {\bibfield  {journal} {\bibinfo  {journal} {ACS photonics}\ }\textbf {\bibinfo {volume} {9}},\ \bibinfo {pages} {1096} (\bibinfo {year} {2022})}\BibitemShut {NoStop}%
\bibitem [{\citenamefont {Mukamel}(2000)}]{mukamel2000}%
  \BibitemOpen
  \bibfield  {author} {\bibinfo {author} {\bibfnamefont {S.}~\bibnamefont {Mukamel}},\ }\bibfield  {title} {\bibinfo {title} {Multidimensional femtosecond correlation spectroscopies of electronic and vibrational excitations},\ }\href {https://doi.org/10.1146/annurev.physchem.51.1.691} {\bibfield  {journal} {\bibinfo  {journal} {Annual review of physical chemistry}\ }\textbf {\bibinfo {volume} {51}},\ \bibinfo {pages} {691} (\bibinfo {year} {2000})}\BibitemShut {NoStop}%
\bibitem [{\citenamefont {Jonas}(2003)}]{jonas2003}%
  \BibitemOpen
  \bibfield  {author} {\bibinfo {author} {\bibfnamefont {D.~M.}\ \bibnamefont {Jonas}},\ }\bibfield  {title} {\bibinfo {title} {Two-dimensional femtosecond spectroscopy},\ }\href {https://doi.org/https://doi.org/10.1146/annurev.physchem.54.011002.103907} {\bibfield  {journal} {\bibinfo  {journal} {Annual Review of Physical Chemistry}\ }\textbf {\bibinfo {volume} {54}},\ \bibinfo {pages} {425} (\bibinfo {year} {2003})}\BibitemShut {NoStop}%
\bibitem [{\citenamefont {Abramavicius}\ \emph {et~al.}(2009)\citenamefont {Abramavicius}, \citenamefont {Palmieri}, \citenamefont {Voronine}, \citenamefont {Sanda},\ and\ \citenamefont {Mukamel}}]{abramavicius2009}%
  \BibitemOpen
  \bibfield  {author} {\bibinfo {author} {\bibfnamefont {D.}~\bibnamefont {Abramavicius}}, \bibinfo {author} {\bibfnamefont {B.}~\bibnamefont {Palmieri}}, \bibinfo {author} {\bibfnamefont {D.~V.}\ \bibnamefont {Voronine}}, \bibinfo {author} {\bibfnamefont {F.}~\bibnamefont {Sanda}},\ and\ \bibinfo {author} {\bibfnamefont {S.}~\bibnamefont {Mukamel}},\ }\bibfield  {title} {\bibinfo {title} {Coherent multidimensional optical spectroscopy of excitons in molecular aggregates; quasiparticle versus supermolecule perspectives},\ }\href {https://doi.org/https://doi.org/10.1021/cr800268n} {\bibfield  {journal} {\bibinfo  {journal} {Chemical reviews}\ }\textbf {\bibinfo {volume} {109}},\ \bibinfo {pages} {2350} (\bibinfo {year} {2009})}\BibitemShut {NoStop}%
\bibitem [{\citenamefont {Fuller}\ and\ \citenamefont {Ogilvie}(2015)}]{fuller2015}%
  \BibitemOpen
  \bibfield  {author} {\bibinfo {author} {\bibfnamefont {F.~D.}\ \bibnamefont {Fuller}}\ and\ \bibinfo {author} {\bibfnamefont {J.~P.}\ \bibnamefont {Ogilvie}},\ }\bibfield  {title} {\bibinfo {title} {Experimental implementations of two-dimensional fourier transform electronic spectroscopy},\ }\href {https://doi.org/https://doi.org/10.1146/annurev-physchem-040513-103623} {\bibfield  {journal} {\bibinfo  {journal} {Annual Review of Physical Chemistry}\ }\textbf {\bibinfo {volume} {66}},\ \bibinfo {pages} {667} (\bibinfo {year} {2015})}\BibitemShut {NoStop}%
\bibitem [{\citenamefont {Saurabh}\ and\ \citenamefont {Mukamel}(2016)}]{saurabh2016}%
  \BibitemOpen
  \bibfield  {author} {\bibinfo {author} {\bibfnamefont {P.}~\bibnamefont {Saurabh}}\ and\ \bibinfo {author} {\bibfnamefont {S.}~\bibnamefont {Mukamel}},\ }\bibfield  {title} {\bibinfo {title} {Two-dimensional infrared spectroscopy of vibrational polaritons of molecules in an optical cavity},\ }\href {https://doi.org/https://doi.org/10.1063/1.4944492} {\bibfield  {journal} {\bibinfo  {journal} {The Journal of chemical physics}\ }\textbf {\bibinfo {volume} {144}},\ \bibinfo {pages} {124115} (\bibinfo {year} {2016})}\BibitemShut {NoStop}%
\bibitem [{\citenamefont {Dorfman}\ and\ \citenamefont {Mukamel}(2018)}]{dorfman2018}%
  \BibitemOpen
  \bibfield  {author} {\bibinfo {author} {\bibfnamefont {K.~E.}\ \bibnamefont {Dorfman}}\ and\ \bibinfo {author} {\bibfnamefont {S.}~\bibnamefont {Mukamel}},\ }\bibfield  {title} {\bibinfo {title} {Multidimensional photon correlation spectroscopy of cavity polaritons},\ }\href {https://doi.org/https://doi.org/10.1073/pnas.171944311} {\bibfield  {journal} {\bibinfo  {journal} {Proceedings of the National Academy of Sciences}\ }\textbf {\bibinfo {volume} {115}},\ \bibinfo {pages} {1451} (\bibinfo {year} {2018})}\BibitemShut {NoStop}%
\bibitem [{\citenamefont {Tiwari}(2021)}]{tiwari2021}%
  \BibitemOpen
  \bibfield  {author} {\bibinfo {author} {\bibfnamefont {V.}~\bibnamefont {Tiwari}},\ }\bibfield  {title} {\bibinfo {title} {{Multidimensional electronic spectroscopy in high-definition—Combining spectral, temporal, and spatial resolutions}},\ }\href {https://doi.org/https://doi.org/10.1063/5.0052234} {\bibfield  {journal} {\bibinfo  {journal} {The Journal of Chemical Physics}\ }\textbf {\bibinfo {volume} {154}},\ \bibinfo {pages} {230901} (\bibinfo {year} {2021})}\BibitemShut {NoStop}%
\bibitem [{\citenamefont {Oliver}(2018)}]{oliver2018}%
  \BibitemOpen
  \bibfield  {author} {\bibinfo {author} {\bibfnamefont {T.~A.~A.}\ \bibnamefont {Oliver}},\ }\bibfield  {title} {\bibinfo {title} {Recent advances in multidimensional ultrafast spectroscopy},\ }\href {https://doi.org/10.1098/rsos.171425} {\bibfield  {journal} {\bibinfo  {journal} {Royal Society Open Science}\ }\textbf {\bibinfo {volume} {5}},\ \bibinfo {pages} {171425} (\bibinfo {year} {2018})}\BibitemShut {NoStop}%
\bibitem [{\citenamefont {Gelzinis}\ \emph {et~al.}(2019)\citenamefont {Gelzinis}, \citenamefont {Augulis}, \citenamefont {Butkus}, \citenamefont {Robert},\ and\ \citenamefont {Valkunas}}]{gelzinis2019two}%
  \BibitemOpen
  \bibfield  {author} {\bibinfo {author} {\bibfnamefont {A.}~\bibnamefont {Gelzinis}}, \bibinfo {author} {\bibfnamefont {R.}~\bibnamefont {Augulis}}, \bibinfo {author} {\bibfnamefont {V.}~\bibnamefont {Butkus}}, \bibinfo {author} {\bibfnamefont {B.}~\bibnamefont {Robert}},\ and\ \bibinfo {author} {\bibfnamefont {L.}~\bibnamefont {Valkunas}},\ }\bibfield  {title} {\bibinfo {title} {Two-dimensional spectroscopy for non-specialists},\ }\href {https://doi.org/10.1016/j.bbabio.2018.12.006} {\bibfield  {journal} {\bibinfo  {journal} {Biochimica et Biophysica Acta (BBA)-Bioenergetics}\ }\textbf {\bibinfo {volume} {1860}},\ \bibinfo {pages} {271} (\bibinfo {year} {2019})}\BibitemShut {NoStop}%
\bibitem [{\citenamefont {F.~Ribeiro}\ \emph {et~al.}(2018)\citenamefont {F.~Ribeiro}, \citenamefont {Dunkelberger}, \citenamefont {Xiang}, \citenamefont {Xiong}, \citenamefont {Simpkins}, \citenamefont {Owrutsky},\ and\ \citenamefont {Yuen-Zhou}}]{ribeiro2018theory}%
  \BibitemOpen
  \bibfield  {author} {\bibinfo {author} {\bibfnamefont {R.}~\bibnamefont {F.~Ribeiro}}, \bibinfo {author} {\bibfnamefont {A.~D.}\ \bibnamefont {Dunkelberger}}, \bibinfo {author} {\bibfnamefont {B.}~\bibnamefont {Xiang}}, \bibinfo {author} {\bibfnamefont {W.}~\bibnamefont {Xiong}}, \bibinfo {author} {\bibfnamefont {B.~S.}\ \bibnamefont {Simpkins}}, \bibinfo {author} {\bibfnamefont {J.~C.}\ \bibnamefont {Owrutsky}},\ and\ \bibinfo {author} {\bibfnamefont {J.}~\bibnamefont {Yuen-Zhou}},\ }\bibfield  {title} {\bibinfo {title} {Theory for nonlinear spectroscopy of vibrational polaritons},\ }\href {https://doi.org/https://doi.org/10.1021/acs.jpclett.8b01176} {\bibfield  {journal} {\bibinfo  {journal} {The journal of physical chemistry letters}\ }\textbf {\bibinfo {volume} {9}},\ \bibinfo {pages} {3766} (\bibinfo {year} {2018})}\BibitemShut {NoStop}%
\bibitem [{\citenamefont {Xiang}\ \emph {et~al.}(2018)\citenamefont {Xiang}, \citenamefont {Ribeiro}, \citenamefont {Dunkelberger}, \citenamefont {Wang}, \citenamefont {Li}, \citenamefont {Simpkins}, \citenamefont {Owrutsky}, \citenamefont {Yuen-Zhou},\ and\ \citenamefont {Xiong}}]{xiang2018two}%
  \BibitemOpen
  \bibfield  {author} {\bibinfo {author} {\bibfnamefont {B.}~\bibnamefont {Xiang}}, \bibinfo {author} {\bibfnamefont {R.~F.}\ \bibnamefont {Ribeiro}}, \bibinfo {author} {\bibfnamefont {A.~D.}\ \bibnamefont {Dunkelberger}}, \bibinfo {author} {\bibfnamefont {J.}~\bibnamefont {Wang}}, \bibinfo {author} {\bibfnamefont {Y.}~\bibnamefont {Li}}, \bibinfo {author} {\bibfnamefont {B.~S.}\ \bibnamefont {Simpkins}}, \bibinfo {author} {\bibfnamefont {J.~C.}\ \bibnamefont {Owrutsky}}, \bibinfo {author} {\bibfnamefont {J.}~\bibnamefont {Yuen-Zhou}},\ and\ \bibinfo {author} {\bibfnamefont {W.}~\bibnamefont {Xiong}},\ }\bibfield  {title} {\bibinfo {title} {Two-dimensional infrared spectroscopy of vibrational polaritons},\ }\href {https://doi.org/https://doi.org/10.1073/pnas.1722063115} {\bibfield  {journal} {\bibinfo  {journal} {Proceedings of the National Academy of Sciences}\ }\textbf {\bibinfo {volume} {115}},\ \bibinfo {pages} {4845} (\bibinfo {year} {2018})}\BibitemShut {NoStop}%
\bibitem [{\citenamefont {Simpkins}\ \emph {et~al.}(2023)\citenamefont {Simpkins}, \citenamefont {Yang}, \citenamefont {Dunkelberger}, \citenamefont {Vurgaftman}, \citenamefont {Owrutsky},\ and\ \citenamefont {Xiong}}]{simpkins2023}%
  \BibitemOpen
  \bibfield  {author} {\bibinfo {author} {\bibfnamefont {B.~S.}\ \bibnamefont {Simpkins}}, \bibinfo {author} {\bibfnamefont {Z.}~\bibnamefont {Yang}}, \bibinfo {author} {\bibfnamefont {A.~D.}\ \bibnamefont {Dunkelberger}}, \bibinfo {author} {\bibfnamefont {I.}~\bibnamefont {Vurgaftman}}, \bibinfo {author} {\bibfnamefont {J.~C.}\ \bibnamefont {Owrutsky}},\ and\ \bibinfo {author} {\bibfnamefont {W.}~\bibnamefont {Xiong}},\ }\bibfield  {title} {\bibinfo {title} {Comment on “isolating polaritonic 2d-ir transmission spectra”},\ }\href {https://doi.org/https://doi.org/10.1021/acs.jpclett.2c01264} {\bibfield  {journal} {\bibinfo  {journal} {The Journal of Physical Chemistry Letters}\ }\textbf {\bibinfo {volume} {14}},\ \bibinfo {pages} {983} (\bibinfo {year} {2023})}\BibitemShut {NoStop}%
\bibitem [{\citenamefont {Li}\ and\ \citenamefont {Cundiff}(2017)}]{Li20172DCoherent}%
  \BibitemOpen
  \bibfield  {author} {\bibinfo {author} {\bibfnamefont {H.}~\bibnamefont {Li}}\ and\ \bibinfo {author} {\bibfnamefont {S.~T.}\ \bibnamefont {Cundiff}},\ }\bibfield  {title} {\bibinfo {title} {{Chapter {{One}} - {{2D Coherent Spectroscopy}} of {{Electronic Transitions}}}},\ }in\ \href {https://doi.org/10.1016/bs.aamop.2017.03.004} {\emph {\bibinfo {booktitle} {Advances {{In Atomic}}, {{Molecular}}, and {{Optical Physics}}}}},\ Vol.~\bibinfo {volume} {66},\ \bibinfo {editor} {edited by\ \bibinfo {editor} {\bibfnamefont {E.}~\bibnamefont {Arimondo}}, \bibinfo {editor} {\bibfnamefont {C.~C.}\ \bibnamefont {Lin}},\ and\ \bibinfo {editor} {\bibfnamefont {S.~F.}\ \bibnamefont {Yelin}}}\ (\bibinfo  {publisher} {{Academic Press}},\ \bibinfo {year} {2017})\ p.~\bibinfo {pages} {1}\BibitemShut {NoStop}%
\bibitem [{\citenamefont {Jaynes}\ and\ \citenamefont {Cummings}(1963)}]{Jaynes1963}%
  \BibitemOpen
  \bibfield  {author} {\bibinfo {author} {\bibfnamefont {E.~T.}\ \bibnamefont {Jaynes}}\ and\ \bibinfo {author} {\bibfnamefont {F.~W.}\ \bibnamefont {Cummings}},\ }\bibfield  {title} {\bibinfo {title} {{Comparison of Quantum and Semiclassical Radiation Theories with Application to the Beam Maser}},\ }\href {https://doi.org/10.1109/PROC.1963.1664} {\bibfield  {journal} {\bibinfo  {journal} {Proc. IEEE}\ }\textbf {\bibinfo {volume} {51}},\ \bibinfo {pages} {89} (\bibinfo {year} {1963})}\BibitemShut {NoStop}%
\bibitem [{\citenamefont {Shore}\ and\ \citenamefont {Knight}(1993)}]{Shore1993}%
  \BibitemOpen
  \bibfield  {author} {\bibinfo {author} {\bibfnamefont {B.~W.}\ \bibnamefont {Shore}}\ and\ \bibinfo {author} {\bibfnamefont {P.~L.}\ \bibnamefont {Knight}},\ }\bibfield  {title} {\bibinfo {title} {{The {{Jaynes-Cummings Model}}}},\ }\href {https://doi.org/10.1080/09500349314551321} {\bibfield  {journal} {\bibinfo  {journal} {J. Mod. Opt.}\ }\textbf {\bibinfo {volume} {40}},\ \bibinfo {pages} {1195} (\bibinfo {year} {1993})}\BibitemShut {NoStop}%
\bibitem [{\citenamefont {Dicke}(1954)}]{Dicke1954}%
  \BibitemOpen
  \bibfield  {author} {\bibinfo {author} {\bibfnamefont {R.}~\bibnamefont {Dicke}},\ }\bibfield  {title} {\bibinfo {title} {{Coherence in {{Spontaneous Radiation Processes}}}},\ }\href {https://doi.org/10.1103/PhysRev.93.99} {\bibfield  {journal} {\bibinfo  {journal} {Phys. Rev.}\ }\textbf {\bibinfo {volume} {93}},\ \bibinfo {pages} {99} (\bibinfo {year} {1954})}\BibitemShut {NoStop}%
\bibitem [{\citenamefont {Tavis}\ and\ \citenamefont {Cummings}(1968)}]{Tavis1968}%
  \BibitemOpen
  \bibfield  {author} {\bibinfo {author} {\bibfnamefont {M.}~\bibnamefont {Tavis}}\ and\ \bibinfo {author} {\bibfnamefont {F.~W.}\ \bibnamefont {Cummings}},\ }\bibfield  {title} {\bibinfo {title} {{Exact Solution for an {{N-molecule-radiation-field Hamiltonian}}}},\ }\href {https://doi.org/10.1103/PhysRev.170.379} {\bibfield  {journal} {\bibinfo  {journal} {Phys. Rev.}\ }\textbf {\bibinfo {volume} {170}},\ \bibinfo {pages} {379} (\bibinfo {year} {1968})}\BibitemShut {NoStop}%
\bibitem [{\citenamefont {Garraway}(2011)}]{garraway2011}%
  \BibitemOpen
  \bibfield  {author} {\bibinfo {author} {\bibfnamefont {B.~M.}\ \bibnamefont {Garraway}},\ }\bibfield  {title} {\bibinfo {title} {The dicke model in quantum optics: Dicke model revisited},\ }\href {https://doi.org/https://doi.org/10.1098/rsta.2010.0333} {\bibfield  {journal} {\bibinfo  {journal} {Philosophical Transactions of the Royal Society A: Mathematical, Physical and Engineering Sciences}\ }\textbf {\bibinfo {volume} {369}},\ \bibinfo {pages} {1137} (\bibinfo {year} {2011})}\BibitemShut {NoStop}%
\bibitem [{\citenamefont {Mewes}\ \emph {et~al.}(2020)\citenamefont {Mewes}, \citenamefont {Wang}, \citenamefont {Ingle}, \citenamefont {B{\"o}rjesson},\ and\ \citenamefont {Chergui}}]{mewes2020}%
  \BibitemOpen
  \bibfield  {author} {\bibinfo {author} {\bibfnamefont {L.}~\bibnamefont {Mewes}}, \bibinfo {author} {\bibfnamefont {M.}~\bibnamefont {Wang}}, \bibinfo {author} {\bibfnamefont {R.~A.}\ \bibnamefont {Ingle}}, \bibinfo {author} {\bibfnamefont {K.}~\bibnamefont {B{\"o}rjesson}},\ and\ \bibinfo {author} {\bibfnamefont {M.}~\bibnamefont {Chergui}},\ }\bibfield  {title} {\bibinfo {title} {Energy relaxation pathways between light-matter states revealed by coherent two-dimensional spectroscopy},\ }\href {https://doi.org/https://doi.org/10.1038/s42005-020-00424-z} {\bibfield  {journal} {\bibinfo  {journal} {Communications Physics}\ }\textbf {\bibinfo {volume} {3}},\ \bibinfo {pages} {1} (\bibinfo {year} {2020})}\BibitemShut {NoStop}%
\bibitem [{\citenamefont {Kasha}(1963)}]{Kasha1963}%
  \BibitemOpen
  \bibfield  {author} {\bibinfo {author} {\bibfnamefont {M.}~\bibnamefont {Kasha}},\ }\bibfield  {title} {\bibinfo {title} {{Energy Transfer Mechanisms and the Molecular Exciton Model for Molecular Aggregates}},\ }\href {https://doi.org/https://doi.org/10.2307/3571331} {\bibfield  {journal} {\bibinfo  {journal} {Radiation Research}\ }\textbf {\bibinfo {volume} {20}},\ \bibinfo {pages} {55} (\bibinfo {year} {1963})}\BibitemShut {NoStop}%
\bibitem [{\citenamefont {Hestand}\ and\ \citenamefont {Spano}(2018)}]{Hestand2018}%
  \BibitemOpen
  \bibfield  {author} {\bibinfo {author} {\bibfnamefont {N.~J.}\ \bibnamefont {Hestand}}\ and\ \bibinfo {author} {\bibfnamefont {F.~C.}\ \bibnamefont {Spano}},\ }\bibfield  {title} {\bibinfo {title} {Expanded theory of h- and j‑molecular aggregates: The effects of vibronic coupling and intermolecular charge transfer},\ }\href {https://doi.org/https://doi.org/10.1021/acs.chemrev.7b00581} {\bibfield  {journal} {\bibinfo  {journal} {Chem. Rev.}\ }\textbf {\bibinfo {volume} {118}},\ \bibinfo {pages} {7069–7163} (\bibinfo {year} {2018})}\BibitemShut {NoStop}%
\bibitem [{\citenamefont {Fassioli}\ \emph {et~al.}(2021)\citenamefont {Fassioli}, \citenamefont {Park}, \citenamefont {Bard},\ and\ \citenamefont {Scholes}}]{fassioli2021}%
  \BibitemOpen
  \bibfield  {author} {\bibinfo {author} {\bibfnamefont {F.}~\bibnamefont {Fassioli}}, \bibinfo {author} {\bibfnamefont {K.~H.}\ \bibnamefont {Park}}, \bibinfo {author} {\bibfnamefont {S.~E.}\ \bibnamefont {Bard}},\ and\ \bibinfo {author} {\bibfnamefont {G.~D.}\ \bibnamefont {Scholes}},\ }\bibfield  {title} {\bibinfo {title} {Femtosecond photophysics of molecular polaritons},\ }\href {https://doi.org/https://doi.org/10.1021/acs.jpclett.1c03183} {\bibfield  {journal} {\bibinfo  {journal} {The Journal of Physical Chemistry Letters}\ }\textbf {\bibinfo {volume} {12}},\ \bibinfo {pages} {11444} (\bibinfo {year} {2021})}\BibitemShut {NoStop}%
\bibitem [{\citenamefont {Zhang}\ \emph {et~al.}(2019)\citenamefont {Zhang}, \citenamefont {Wang}, \citenamefont {Yi}, \citenamefont {Zubairy}, \citenamefont {Scully},\ and\ \citenamefont {Mukamel}}]{Zhang2019}%
  \BibitemOpen
  \bibfield  {author} {\bibinfo {author} {\bibfnamefont {Z.}~\bibnamefont {Zhang}}, \bibinfo {author} {\bibfnamefont {K.}~\bibnamefont {Wang}}, \bibinfo {author} {\bibfnamefont {Z.}~\bibnamefont {Yi}}, \bibinfo {author} {\bibfnamefont {M.~S.}\ \bibnamefont {Zubairy}}, \bibinfo {author} {\bibfnamefont {M.~O.}\ \bibnamefont {Scully}},\ and\ \bibinfo {author} {\bibfnamefont {S.}~\bibnamefont {Mukamel}},\ }\bibfield  {title} {\bibinfo {title} {{Polariton-Assisted Cooperativity of Molecules in Microcavities Monitored by Two-Dimensional Infrared Spectroscopy }},\ }\href {https://doi.org/https://doi.org/10.1021/acs.jpclett.9b00979} {\bibfield  {journal} {\bibinfo  {journal} {J. Phys. Chem. Lett.}\ }\textbf {\bibinfo {volume} {10}},\ \bibinfo {pages} {4448–4454} (\bibinfo {year} {2019})}\BibitemShut {NoStop}%
\bibitem [{\citenamefont {Mondal}\ \emph {et~al.}(2023)\citenamefont {Mondal}, \citenamefont {Koessler}, \citenamefont {Provazza}, \citenamefont {Vamivakas}, \citenamefont {Cundiff}, \citenamefont {Krauss},\ and\ \citenamefont {Huo}}]{Mondal2023}%
  \BibitemOpen
  \bibfield  {author} {\bibinfo {author} {\bibfnamefont {M.~E.}\ \bibnamefont {Mondal}}, \bibinfo {author} {\bibfnamefont {E.~R.}\ \bibnamefont {Koessler}}, \bibinfo {author} {\bibfnamefont {J.}~\bibnamefont {Provazza}}, \bibinfo {author} {\bibfnamefont {A.~N.}\ \bibnamefont {Vamivakas}}, \bibinfo {author} {\bibfnamefont {S.~T.}\ \bibnamefont {Cundiff}}, \bibinfo {author} {\bibfnamefont {T.~D.}\ \bibnamefont {Krauss}},\ and\ \bibinfo {author} {\bibfnamefont {P.}~\bibnamefont {Huo}},\ }\bibfield  {title} {\bibinfo {title} {{Quantum dynamics simulations of the 2D spectroscopy for exciton polaritons}},\ }\href {https://doi.org/https://doi.org/10.1063/5.0166188} {\bibfield  {journal} {\bibinfo  {journal} {The Journal of Chemical Physics}\ }\textbf {\bibinfo {volume} {159}},\ \bibinfo {pages} {094102} (\bibinfo {year} {2023})}\BibitemShut {NoStop}%
\bibitem [{\citenamefont {Zhang}\ \emph {et~al.}(2023)\citenamefont {Zhang}, \citenamefont {Nie}, \citenamefont {Lei},\ and\ \citenamefont {Mukamel}}]{Zhang2023}%
  \BibitemOpen
  \bibfield  {author} {\bibinfo {author} {\bibfnamefont {Z.}~\bibnamefont {Zhang}}, \bibinfo {author} {\bibfnamefont {X.}~\bibnamefont {Nie}}, \bibinfo {author} {\bibfnamefont {D.}~\bibnamefont {Lei}},\ and\ \bibinfo {author} {\bibfnamefont {S.}~\bibnamefont {Mukamel}},\ }\bibfield  {title} {\bibinfo {title} {Multidimensional coherent spectroscopy of molecular polaritons: Langevin approach},\ }\href {https://doi.org/https://doi.org/10.1103/PhysRevLett.130.103001} {\bibfield  {journal} {\bibinfo  {journal} {Phys. Rev. Lett.}\ }\textbf {\bibinfo {volume} {130}},\ \bibinfo {pages} {103001} (\bibinfo {year} {2023})}\BibitemShut {NoStop}%
\bibitem [{\citenamefont {{S{\'a}nchez-Barquilla}}\ \emph {et~al.}(2022)\citenamefont {{S{\'a}nchez-Barquilla}}, \citenamefont {{Fern{\'a}ndez-Dom{\'i}nguez}}, \citenamefont {Feist},\ and\ \citenamefont {{Garc{\'i}a-Vidal}}}]{Sanchez-Barquilla2022Perspective}%
  \BibitemOpen
  \bibfield  {author} {\bibinfo {author} {\bibfnamefont {M.}~\bibnamefont {{S{\'a}nchez-Barquilla}}}, \bibinfo {author} {\bibfnamefont {A.~I.}\ \bibnamefont {{Fern{\'a}ndez-Dom{\'i}nguez}}}, \bibinfo {author} {\bibfnamefont {J.}~\bibnamefont {Feist}},\ and\ \bibinfo {author} {\bibfnamefont {F.~J.}\ \bibnamefont {{Garc{\'i}a-Vidal}}},\ }\bibfield  {title} {\bibinfo {title} {{A {{Theoretical Perspective}} on {{Molecular Polaritonics}}}},\ }\href {https://doi.org/10.1021/acsphotonics.2c00048} {\bibfield  {journal} {\bibinfo  {journal} {ACS Photonics}\ }\textbf {\bibinfo {volume} {9}},\ \bibinfo {pages} {1830} (\bibinfo {year} {2022})}\BibitemShut {NoStop}%
\bibitem [{\citenamefont {Breuer}\ \emph {et~al.}(2002)\citenamefont {Breuer}, \citenamefont {Petruccione} \emph {et~al.}}]{breuer2002}%
  \BibitemOpen
  \bibfield  {author} {\bibinfo {author} {\bibfnamefont {H.-P.}\ \bibnamefont {Breuer}}, \bibinfo {author} {\bibfnamefont {F.}~\bibnamefont {Petruccione}}, \emph {et~al.},\ }\href@noop {} {\emph {\bibinfo {title} {The theory of open quantum systems}}}\ (\bibinfo  {publisher} {Oxford University Press on Demand},\ \bibinfo {year} {2002})\BibitemShut {NoStop}%
\bibitem [{\citenamefont {Jeske}\ and\ \citenamefont {Cole}(2013)}]{jeske2013}%
  \BibitemOpen
  \bibfield  {author} {\bibinfo {author} {\bibfnamefont {J.}~\bibnamefont {Jeske}}\ and\ \bibinfo {author} {\bibfnamefont {J.~H.}\ \bibnamefont {Cole}},\ }\bibfield  {title} {\bibinfo {title} {Derivation of markovian master equations for spatially correlated decoherence},\ }\href {https://doi.org/https://doi.org/10.1103/PhysRevA.87.052138} {\bibfield  {journal} {\bibinfo  {journal} {Physical Review A}\ }\textbf {\bibinfo {volume} {87}},\ \bibinfo {pages} {052138} (\bibinfo {year} {2013})}\BibitemShut {NoStop}%
\bibitem [{\citenamefont {Manzano}(2020)}]{manzano2020}%
  \BibitemOpen
  \bibfield  {author} {\bibinfo {author} {\bibfnamefont {D.}~\bibnamefont {Manzano}},\ }\bibfield  {title} {\bibinfo {title} {A short introduction to the lindblad master equation},\ }\href {https://doi.org/https://doi.org/10.1063/1.5115323} {\bibfield  {journal} {\bibinfo  {journal} {AIP Advances}\ }\textbf {\bibinfo {volume} {10}},\ \bibinfo {pages} {025106} (\bibinfo {year} {2020})}\BibitemShut {NoStop}%
\bibitem [{\citenamefont {Jeske}\ \emph {et~al.}(2015)\citenamefont {Jeske}, \citenamefont {Ing}, \citenamefont {Plenio}, \citenamefont {Huelga},\ and\ \citenamefont {Cole}}]{Jeske2015Bloch}%
  \BibitemOpen
  \bibfield  {author} {\bibinfo {author} {\bibfnamefont {J.}~\bibnamefont {Jeske}}, \bibinfo {author} {\bibfnamefont {D.~J.}\ \bibnamefont {Ing}}, \bibinfo {author} {\bibfnamefont {M.~B.}\ \bibnamefont {Plenio}}, \bibinfo {author} {\bibfnamefont {S.~F.}\ \bibnamefont {Huelga}},\ and\ \bibinfo {author} {\bibfnamefont {J.~H.}\ \bibnamefont {Cole}},\ }\bibfield  {title} {\bibinfo {title} {{Bloch-Redfield equations for modeling light-harvesting complexes}},\ }\href {https://doi.org/10.1063/1.4907370} {\bibfield  {journal} {\bibinfo  {journal} {The Journal of Chemical Physics}\ }\textbf {\bibinfo {volume} {142}},\ \bibinfo {pages} {064104} (\bibinfo {year} {2015})}\BibitemShut {NoStop}%
\bibitem [{\citenamefont {del Pino}\ \emph {et~al.}(2015)\citenamefont {del Pino}, \citenamefont {Feist},\ and\ \citenamefont {Garcia-Vidal}}]{delPino2015}%
  \BibitemOpen
  \bibfield  {author} {\bibinfo {author} {\bibfnamefont {J.}~\bibnamefont {del Pino}}, \bibinfo {author} {\bibfnamefont {J.}~\bibnamefont {Feist}},\ and\ \bibinfo {author} {\bibfnamefont {F.~J.}\ \bibnamefont {Garcia-Vidal}},\ }\bibfield  {title} {\bibinfo {title} {Quantum theory of collective strong coupling of molecular vibrations with a microcavity mode},\ }\href {https://doi.org/10.1088/1367-2630/17/5/053040} {\bibfield  {journal} {\bibinfo  {journal} {New Journal of Physics}\ }\textbf {\bibinfo {volume} {17}},\ \bibinfo {pages} {053040} (\bibinfo {year} {2015})}\BibitemShut {NoStop}%
\bibitem [{\citenamefont {Lentrodt}\ and\ \citenamefont {Evers}(2020)}]{Lentrodt2020}%
  \BibitemOpen
  \bibfield  {author} {\bibinfo {author} {\bibfnamefont {D.}~\bibnamefont {Lentrodt}}\ and\ \bibinfo {author} {\bibfnamefont {J.}~\bibnamefont {Evers}},\ }\bibfield  {title} {\bibinfo {title} {Ab initio few-mode theory for quantum potential scattering problems},\ }\href {https://doi.org/https://doi.org/10.1103/PhysRevX.10.011008} {\bibfield  {journal} {\bibinfo  {journal} {Phys. Rev. X}\ }\textbf {\bibinfo {volume} {10}},\ \bibinfo {pages} {011008} (\bibinfo {year} {2020})}\BibitemShut {NoStop}%
\bibitem [{\citenamefont {Mukamel}(1999)}]{mukamel1999}%
  \BibitemOpen
  \bibfield  {author} {\bibinfo {author} {\bibfnamefont {S.}~\bibnamefont {Mukamel}},\ }\href@noop {} {\emph {\bibinfo {title} {Principles of nonlinear optical spectroscopy}}}\ (\bibinfo  {publisher} {Oxford University Press},\ \bibinfo {year} {1999})\BibitemShut {NoStop}%
\bibitem [{\citenamefont {Cho}(2009)}]{cho2009}%
  \BibitemOpen
  \bibfield  {author} {\bibinfo {author} {\bibfnamefont {M.}~\bibnamefont {Cho}},\ }\href@noop {} {\emph {\bibinfo {title} {Two-dimensional optical spectroscopy}}}\ (\bibinfo  {publisher} {CRC press},\ \bibinfo {year} {2009})\BibitemShut {NoStop}%
\bibitem [{\citenamefont {Hamm}\ and\ \citenamefont {Zanni}(2011)}]{hamm2011}%
  \BibitemOpen
  \bibfield  {author} {\bibinfo {author} {\bibfnamefont {P.}~\bibnamefont {Hamm}}\ and\ \bibinfo {author} {\bibfnamefont {M.}~\bibnamefont {Zanni}},\ }\href@noop {} {\emph {\bibinfo {title} {Concepts and methods of 2D infrared spectroscopy}}}\ (\bibinfo  {publisher} {Cambridge University Press},\ \bibinfo {year} {2011})\BibitemShut {NoStop}%
\bibitem [{\citenamefont {Gelin}\ \emph {et~al.}(2009)\citenamefont {Gelin}, \citenamefont {Egorova},\ and\ \citenamefont {Domcke}}]{gelin2009}%
  \BibitemOpen
  \bibfield  {author} {\bibinfo {author} {\bibfnamefont {M.~F.}\ \bibnamefont {Gelin}}, \bibinfo {author} {\bibfnamefont {D.}~\bibnamefont {Egorova}},\ and\ \bibinfo {author} {\bibfnamefont {W.}~\bibnamefont {Domcke}},\ }\bibfield  {title} {\bibinfo {title} {Efficient calculation of time-and frequency-resolved four-wave-mixing signals},\ }\href {https://doi.org/https://doi.org/10.1021/ar900045d} {\bibfield  {journal} {\bibinfo  {journal} {Accounts of chemical research}\ }\textbf {\bibinfo {volume} {42}},\ \bibinfo {pages} {1290} (\bibinfo {year} {2009})}\BibitemShut {NoStop}%
\bibitem [{\citenamefont {De~Sio}\ \emph {et~al.}(2019)\citenamefont {De~Sio}, \citenamefont {Nguyen},\ and\ \citenamefont {Lienau}}]{deSio2019}%
  \BibitemOpen
  \bibfield  {author} {\bibinfo {author} {\bibfnamefont {A.}~\bibnamefont {De~Sio}}, \bibinfo {author} {\bibfnamefont {X.~T.}\ \bibnamefont {Nguyen}},\ and\ \bibinfo {author} {\bibfnamefont {C.}~\bibnamefont {Lienau}},\ }\bibfield  {title} {\bibinfo {title} {Signatures of strong vibronic coupling mediating coherent charge transfer in two-dimensional electronic spectroscopy},\ }\href {https://doi.org/https://doi.org/10.1515/zna-2019-0150} {\bibfield  {journal} {\bibinfo  {journal} {Zeitschrift f{\"u}r Naturforschung A}\ }\textbf {\bibinfo {volume} {74}},\ \bibinfo {pages} {721} (\bibinfo {year} {2019})}\BibitemShut {NoStop}%
\bibitem [{\citenamefont {Valkunas}\ \emph {et~al.}(2013)\citenamefont {Valkunas}, \citenamefont {Abramavicious},\ and\ \citenamefont {Man{\v c}al}}]{valkunas2013}%
  \BibitemOpen
  \bibfield  {author} {\bibinfo {author} {\bibfnamefont {L.}~\bibnamefont {Valkunas}}, \bibinfo {author} {\bibfnamefont {D.}~\bibnamefont {Abramavicious}},\ and\ \bibinfo {author} {\bibfnamefont {T.}~\bibnamefont {Man{\v c}al}},\ }\href@noop {} {\emph {\bibinfo {title} {Molecular Excitation Dynamics}}}\ (\bibinfo  {publisher} {Wiley},\ \bibinfo {year} {2013})\BibitemShut {NoStop}%
\bibitem [{\citenamefont {Johansson}\ \emph {et~al.}(2013)\citenamefont {Johansson}, \citenamefont {Nation},\ and\ \citenamefont {Nori}}]{johansson2013}%
  \BibitemOpen
  \bibfield  {author} {\bibinfo {author} {\bibfnamefont {J.}~\bibnamefont {Johansson}}, \bibinfo {author} {\bibfnamefont {P.}~\bibnamefont {Nation}},\ and\ \bibinfo {author} {\bibfnamefont {F.}~\bibnamefont {Nori}},\ }\bibfield  {title} {\bibinfo {title} {{QuTiP} 2: A python framework for the dynamics of open quantum systems},\ }\href {https://doi.org/10.1016/j.cpc.2012.11.019} {\bibfield  {journal} {\bibinfo  {journal} {Computer Physics Communications}\ }\textbf {\bibinfo {volume} {184}},\ \bibinfo {pages} {1234} (\bibinfo {year} {2013})}\BibitemShut {NoStop}%
\bibitem [{\citenamefont {Neuman}\ and\ \citenamefont {Aizpurua}(2018)}]{neuman2018}%
  \BibitemOpen
  \bibfield  {author} {\bibinfo {author} {\bibfnamefont {T.}~\bibnamefont {Neuman}}\ and\ \bibinfo {author} {\bibfnamefont {J.}~\bibnamefont {Aizpurua}},\ }\bibfield  {title} {\bibinfo {title} {Origin of the asymmetric light emission from molecular exciton--polaritons},\ }\href {https://doi.org/https://doi.org/10.1364/OPTICA.5.001247} {\bibfield  {journal} {\bibinfo  {journal} {Optica}\ }\textbf {\bibinfo {volume} {5}},\ \bibinfo {pages} {1247} (\bibinfo {year} {2018})}\BibitemShut {NoStop}%
\bibitem [{\citenamefont {Takemura}\ \emph {et~al.}(2015)\citenamefont {Takemura}, \citenamefont {Trebaol}, \citenamefont {Anderson}, \citenamefont {Kohnle}, \citenamefont {L\'eger}, \citenamefont {Oberli}, \citenamefont {Portella-Oberli},\ and\ \citenamefont {Deveaud}}]{takemura2015}%
  \BibitemOpen
  \bibfield  {author} {\bibinfo {author} {\bibfnamefont {N.}~\bibnamefont {Takemura}}, \bibinfo {author} {\bibfnamefont {S.}~\bibnamefont {Trebaol}}, \bibinfo {author} {\bibfnamefont {M.~D.}\ \bibnamefont {Anderson}}, \bibinfo {author} {\bibfnamefont {V.}~\bibnamefont {Kohnle}}, \bibinfo {author} {\bibfnamefont {Y.}~\bibnamefont {L\'eger}}, \bibinfo {author} {\bibfnamefont {D.~Y.}\ \bibnamefont {Oberli}}, \bibinfo {author} {\bibfnamefont {M.~T.}\ \bibnamefont {Portella-Oberli}},\ and\ \bibinfo {author} {\bibfnamefont {B.}~\bibnamefont {Deveaud}},\ }\bibfield  {title} {\bibinfo {title} {Two-dimensional fourier transform spectroscopy of exciton-polaritons and their interactions},\ }\href {https://doi.org/https://doi.org/10.1103/PhysRevB.92.125415} {\bibfield  {journal} {\bibinfo  {journal} {Phys. Rev. B}\ }\textbf {\bibinfo {volume} {92}},\ \bibinfo {pages} {125415} (\bibinfo {year} {2015})}\BibitemShut {NoStop}%
\bibitem [{\citenamefont {Finkelstein-Shapiro}\ \emph {et~al.}(2021)\citenamefont {Finkelstein-Shapiro}, \citenamefont {Mante}, \citenamefont {Sarisozen}, \citenamefont {Wittenbecher}, \citenamefont {Minda}, \citenamefont {Balci}, \citenamefont {Pullerits},\ and\ \citenamefont {Zigmantas}}]{finkelstein2021}%
  \BibitemOpen
  \bibfield  {author} {\bibinfo {author} {\bibfnamefont {D.}~\bibnamefont {Finkelstein-Shapiro}}, \bibinfo {author} {\bibfnamefont {P.-A.}\ \bibnamefont {Mante}}, \bibinfo {author} {\bibfnamefont {S.}~\bibnamefont {Sarisozen}}, \bibinfo {author} {\bibfnamefont {L.}~\bibnamefont {Wittenbecher}}, \bibinfo {author} {\bibfnamefont {I.}~\bibnamefont {Minda}}, \bibinfo {author} {\bibfnamefont {S.}~\bibnamefont {Balci}}, \bibinfo {author} {\bibfnamefont {T.}~\bibnamefont {Pullerits}},\ and\ \bibinfo {author} {\bibfnamefont {D.}~\bibnamefont {Zigmantas}},\ }\bibfield  {title} {\bibinfo {title} {Understanding radiative transitions and relaxation pathways in plexcitons},\ }\href {https://doi.org/10.1016/j.chempr.2021.02.028} {\bibfield  {journal} {\bibinfo  {journal} {Chem}\ }\textbf {\bibinfo {volume} {7}},\ \bibinfo {pages} {1092} (\bibinfo {year} {2021})}\BibitemShut {NoStop}%
\end{thebibliography}%

\end{document}